\documentclass[11pt,letterpaper]{article}
\usepackage{geometry}
\usepackage{tikz}
\usetikzlibrary{arrows.meta,bending}
\usepackage{graphicx} 
\usepackage[utf8]{inputenc}
\usepackage{amsmath}
\usepackage{amsfonts}
\usepackage{latexsym}
\usepackage{amsthm}
\usepackage{comment}
\usepackage{float}
\usepackage[colorlinks=true, allcolors=blue]{hyperref}
\usepackage{authblk}
\author[1]{Kabir Narayanan}
\author[2]{Abigail Perryman}
\affil[1]{Universit\'e Paris-Saclay}
\affil[2]{University of Texas, Austin}
\affil[3]{Department of Mathematics, Rutgers University (New Brunswick)}
\title{Asymptotic Momentum of Dirac Particles\\ in One Space Dimension}
\author[3]{A. Shadi Tahvildar-Zadeh}
\date{January 2026}

\newtheorem{theorem}{Theorem}[section]
\newtheorem{corollary}{Corollary}[theorem]
\newtheorem{lemma}[theorem]{Lemma}

\newtheorem{definition}[theorem]{Definition}
\newtheorem{remark}{Remark}

\newcommand{\edit}[1]{\textcolor{black}{#1}}

\begin{document}

\maketitle
\begin{abstract}
   We analyze the trajectories of a massive particle in one space dimension whose motion is guided by a spin-half wave function that evolves according to the free Dirac equation, with its initial wave function being a Gaussian wave packet with a nonzero expected value of momentum $k$\edit{.} 
   We prove that at large times, the  wave function \edit{is approximately equal to the superposition of two wave packets traveling in opposite directions}
   , which \edit{results in} trajectories with \edit{approximately constant} asymptotic momentum $k$ and asymptotic energy \edit{$\pm c^2\sqrt{m^2+k^2}$, with $m$ the rest mass of the particle and $c$ the speed of light}.  The sign of the asymptotic energy is determined by the initial position of the particle. Particles with negative energy will have an asymptotic velocity that is in the opposite direction of their momentum.
  The proof uses the stationary phase approximation method, for which we establish a rigorous error bound.
\end{abstract}
\section{Introduction}
In his 1923 paper ``A Quantum Theory of the Scattering of X-Rays by Light Elements," Arthur Compton successfully explained the scattering of high-energy ``light quanta" in an X-ray beam off electrons in a graphite plate by combining results of the (at the time) new quantum theory with the old classical laws of energy and momentum conservation in particle mechanics. Compton observed that the \edit{frequency of the }rays reflected from the graphite plate had\edit{, in addition to the peak predicted by classical electrodynamics at the incident frequency, a second one at} a lower frequency than \edit{the incident}, and using the Einstein--deBroglie relations for the quantum of light's energy and momentum, $E = \hbar\omega$, $\textbf{p} = \hbar \textbf{k}$ (with $\omega = c|\textbf{k}|$), he found that the shift in frequency corresponded exactly to the energy of the scattered electron.  He gave a formula for the associated scattering angle that he then showed matched his experimental results very well. This work won him a Nobel Prize in 1927, and led to more general acceptance of Einstein's particle theory of light.

While ``photon" was not yet in Compton's vocabulary, he writes that in contrast to the classical wave theory of light, ``we may suppose that any particular quantum of X-rays is not scattered by all the electrons in the radiator, but spends all of its energy upon some particular electron" \cite{PhysRev.21.483}. Thus coherent light rays in Compton's theory behave much like classical billiard balls, and indeed Compton makes use of classical conservation principles of particle mechanics---which are by no means fundamental in interacting quantum systems---to predict the behavior of such particles. By contrast, in orthodox quantum mechanics \edit{particles do not have well-defined positions} until they are measured, and \edit{one }can only make statistical predictions about the outcomes of measurements (either by taking the expected values of the {\em observables} (self-adjoint operators) involved, or by postulating collapse rules). Modern renditions of Compton scattering require Quantum Field Theory, in which fields are fundamental, not particles, since the latter can be created and annihilated during the scattering process.

Yet in \cite{Kiessling_2020} Kiessling {\em et al.} demonstrated that an interacting quantum-mechanical model of a fixed number of electron and photon  {\em particles} in one space dimension is possible via a Lorentz-covariant generalization of Bohmian mechanics known as the ``Hypersurface Bohm-Dirac theory" \cite{DGMZ99}. Numerical experiments performed in \cite{Kiessling_2020} exhibit \edit{trajectories of }photon and electron which bounce off one another, akin to Compton scattering.  \edit{Questions }however remain regarding the relationship between the quantum mechanical notions of energy and momentum associated with operators acting on the Hilbert space of wave functions, and their definite Bohmian counterparts, which are well-defined for a given massive particle's trajectory. 

In this context several questions can be asked: to what extent does the relativistic momentum and energy along a free electron's trajectory align with the expected values of the corresponding operators? Can we meaningfully ascribe an `outgoing' momentum to a particle despite entanglement effects? Can Compton's plane wave assumption (used to derive Compton's formula) be justified using a Bohmian formulation  of the scattering process?

In this paper we address the first of these questions by examining the trajectories of a single free electron, and prove that at large times, the electron wave function becomes \edit{effectively the superposition of two wave packets traveling in opposite directions}, at least in the case the initial wave function is a Gaussian wave packet with a non-zero expected value of momentum $k$\edit{.}  The freely evolving wave function then guides the particle to eventually travel with \edit{approximately} constant velocity corresponding to \edit{asymptotic} momentum  $k_{\mbox{\tiny asym}}= k$ and energy $E_{\mbox{\tiny asym}} = \pm \edit{c^2\sqrt{m^2 + k^2}}$. The sign of the asymptotic energy is determined by the initial position of the particle, and negative energy particles acquire velocities that are in the opposite direction of their momenta.
\subsection{Nonrelativistic particles}
Before we delve into the relativistic realm, we briefly review the nonrelativistic case of particles whose wave function satisfies the free Schr\"odinger equation (see Holland \cite{HollandBOOK} for details):
\begin{equation}\label{schrod}
    i \partial_t \psi = - \frac{1}{2} \partial^2_s \psi,
\end{equation}
where \edit{$s$ denotes the position variable and }we have set $\edit{m} = \hbar = 1$. As initial data we take a Gaussian wave packet \edit{in position space }with mean $0$, standard deviation $\sigma_{\mbox{\tiny pos}}= \frac{1}{2}$, and \edit{mean }frequency $k_0 \in \mathbb{R}$:  
\begin{equation}\label{initSchrod}
    \psi(0,s) = \sqrt[4]{\frac{2}{\pi}}e^{-s^2 + ik_0 s} =: \psi_0(s).
\end{equation}
The Fourier transform of the data will be
\begin{equation}\label{initSchrodmom}
    \hat{\psi}_0(k) = \frac{1}{\sqrt[4]{2\pi}}e^{-(k-k_0)^2/4},
\end{equation}
and thus $|\hat{\psi}_0|^2$, the probability density function of the momentum distribution, will also be a Gaussian, with mean $k_0$ and standard deviation
$\sigma_{\mbox{\tiny mom}} = 1$.  

Since the free Schr\"odinger flow preserves the Gaussian profile, the solution at all later times will still be a Gaussian (with a complex variance), 
which gives rise to the evolving probability density function
\begin{equation}
    \rho(t,s) = |\psi|^2(t,s) = \frac{\sqrt{2/\pi}}{\sqrt{1+4t^2}}e^{-2(s-k_0 t)^2/(1+4t^2)}.
\end{equation}
This is a moving and spreading Gaussian with mean $k_0 t$ and standard deviation $\frac{1}{2}\sqrt{1+4t^2}$. The packet transports to the right if $k_0>0$ or  to the left if $k_0<0$.   The Bohmian velocity field for the particle is
\begin{equation}
    v^{(\psi)}(t,s) = \frac{\mbox{Im}(\psi^*\partial_s \psi)}{\rho} = k_0 + \frac{4t(s-k_0t)}{1+4t^2}.
\end{equation}
Solving the deBroglie-Bohm guiding equation
\begin{equation}
    \frac{dq}{dt} = v^{(\psi)}(t,q(t)),\qquad q(0) = q_0,
\end{equation}
where the initial position $q_0$ is assumed to be randomly distributed according to the initial Gaussian $\rho_0$ \edit{with mean $0$ and standard deviation $\frac{1}{2}$}, thus yields the trajectories\footnote{It is instructive to note that the particle is accelerating even though there is no apparent ``force" on it.  This illustrates that \edit{in }quantum \edit{mechanics }particles obey a different law of motion from classical \edit{mechanics}.} 
\begin{equation}
    q(t) = k_0 t + q_0 \sqrt{1+4t^2}.
\end{equation}
It is thus clear that for large times the asymptotic velocity of the Bohmian particle will be 
\begin{equation}
    \dot{q}(t) \sim k_0 + 2q_0.
\end{equation}
\edit{Assuming that the relationship between momentum and velocity of a particle in nonrelativistic quantum mechanics is the same as in classical mechanics\footnote{\edit{This assumption would of course need to be justified, and as we shall see later in this paper, is in fact not true for {\em relativistic} quantum mechanics. Nevertheless, the agreement shown in this example between the two notions is indicative of their close relationship.}},}
the asymptotic momentum of the particle \edit{would therefore be} a constant that is distributed normally with mean $k_0$, i.e. the same as the expected value of the momentum operator $P = -i \partial_s$, and with variance 1 (which is the same as that of the momentum distribution).  Thus in the non-relativistic case we find perfect agreement between the asymptotic momentum of the particle and the expected value of the momentum operator.  \edit{(See \cite{RDM05} for a much more general result in this direction.)}
\section{Free Dirac Particles}
According to Dirac \cite{Dirac}, a massive spin-half particle such as the electron has a wave function:
\begin{equation}
\psi(t,s) = \begin{pmatrix} \psi_{-}(t,s) \\ \psi_{+}(t,s) \end{pmatrix} \in \mathbb{C}^2
\end{equation}
satisfying the massive Dirac equation:
\begin{equation}\label{eq:DirEl}
-i\edit{c}\hbar \gamma^{\mu} \frac{\partial}{\partial x^{\mu}}\psi + \edit{mc^2}\psi = 0  \hspace{0.4 cm} (\mu = 0, 1).\end{equation}
Here, $t\ \edit{=x^0}$ and $s\ \edit{=x^1}$ represent the generic configuration spacetime variables, we are using Einstein's summation convention where repeated indices are summed over their range (in this case $\mu = 0,1$), \edit{$m$ is the rest mass of the particle, $\hbar$ is Planck's constant, and $c$ is} the speed of light \edit{(which we henceforth set equal to one.)} 
We may use the following representation for the Dirac matrices $\gamma^\mu$:
\begin{equation}
  \gamma^0 = \begin{bmatrix}
0 & 1 \\
1 & 0 \\
\end{bmatrix}, \hspace{0.4 cm} \gamma^1 = \begin{bmatrix}
0 & -1 \\
1 & 0 \\
\end{bmatrix}.
\end{equation}
Equation \eqref{eq:DirEl} needs to be supplemented with {\em initial data} for the wave function at $t=0$:
\begin{equation}\label{initdat}
\psi(0,s) = \mathring{\psi}(s) = \begin{pmatrix}
    \mathring{\psi}_-(s)\\
    \mathring{\psi}_+(s)
\end{pmatrix}.
\end{equation}
The equations for components $\psi_+,\psi_-$ can be decoupled to yield two Klein Gordon equations (one for each component). These can be solved exactly, and we find:
%
\begin{align}\label{psipmsol}
    \psi_{\pm}(t,s) &= \overset{\circ}{\psi_{\pm}}(s \pm t) \nonumber\\ &- \frac{\omega }{2}\int_{s-t}^{s+t} J_1\left(\omega\sqrt{t^2 - (s - \sigma)^2}\right)\frac{\sqrt{t \mp (s- \sigma)}}{\sqrt{t \pm (s- \sigma)}}\overset{\circ}{\psi_{\pm}}(\sigma) \,d\sigma\nonumber \\ &- \frac{i\omega}{2} \int_{s-t}^{s+t} J_0\left(\omega\sqrt{t^2 - (s - \sigma)^2}\right) \overset{\circ}{\psi_{\mp}}(\sigma) \,d\sigma\edit{,}
\end{align}
where we have set $$\omega := \frac{mc}{\hbar}.$$
As is apparent from this form of the solution, each component gets transported in one direction with the velocity of light while developing a tail that gets smeared everywhere inside the light cone, thereby propagating also in the other direction. The smearing is proportional to the mass of the particle.

The components of the {\em Dirac current} $j^\mu := \overline{\psi}\gamma^\mu \psi$ are denoted
\begin{equation}\label{eq:DiracCurrent}\rho := j^{0} = |\psi_{-}|^2 + |\psi_{+}|^2, \hspace{.7cm} J := j^{1} = |\psi_{-}|^2 - |\psi_{+}|^2,\end{equation} 
which together satisfy the {\em continuity equation}
\begin{equation}
    \partial_t \rho + \partial_s J = 0.
\end{equation}
We define the electron {\em velocity field} to be
\begin{equation}\label{def:vel}
 v^{(\psi)}(t,s) := \frac{J(t,s)}{\rho(t,s)} = \frac{|\psi_{-}|^2 - |\psi_{+}|^2}{|\psi_{-}|^2 + |\psi_{+}|^2}(t,s).
\end{equation}
The {\em Guiding Equation} for the electron is then
\begin{equation}\label{eq:guiding}
\frac{dQ}{dt} =  v^{(\psi)}(t,Q(t)).
\end{equation}
As we shall see, the fact that an arbitrary solution of \eqref{eq:DirEl} is a superposition of positive and negative energy states, each  of which propagates in both directions, introduces rapid oscillatory motion for the trajectories ---the famous \textit{Zitterbewegung} effect. 

\subsection{Cayley-Klein parameters and the Bloch sphere representation}

The spinor $\psi \in \mathbb{C}^2$ has four real degrees of freedom. It can be decomposed in the following form, known as the {\em Cayley-Klein parametrization}:

\begin{equation}\label{CK}
\psi = \begin{pmatrix}
\psi_{-} \\
\psi_{+}
\end{pmatrix} = R e^{i\Phi/2} \begin{pmatrix}
\cos{\frac{\Theta}{2} e^{i \Omega/2}} \\
\sin{\frac{\Theta}{2} e^{-i \Omega/2}}
\end{pmatrix} \in \mathbb{C}^2.\end{equation}

Thus $\psi$ has an alternative description in terms of the four real parameters $(R, \Theta, \Omega, \Phi)$, each a function of $t$ and $s$. There is a connection between these and unitary representations of the orthogonal group:
In general, we can represent an element of the {\em special unitary group} $SU(2)$ by two complex numbers \edit{$a_\pm$} such that \edit{$|a_-|^2 + |a_+|^2 = 1$}. $SU(2)$ maps surjectively (it is a double cover) onto the orthogonal group of rotations of $\mathbb{R}^3$, namely $SO(3)$. 
This fact can be used to associate the 2-component spinor $\psi$ with a unit vector in $\mathbb{R}^3$:
\begin{equation}
\vec{n}(\psi) = \frac{1}{R^2} \begin{pmatrix} 2\ \mbox{Re}(\psi_-\psi_+^*) \\ 2\ \mbox{Im}(\psi_-\psi_+^*) \\ |\psi_-|^2 - |\psi_+|^2 \end{pmatrix}.
\end{equation}
Plugging in it the expressions for $\psi_\pm$ from \eqref{CK}, we obtain a familiar expression for a unit 3-vector in terms of two of the Cayley-Klein parameters:

\begin{equation}\label{def:n}
\vec{n}(\psi) = \begin{pmatrix} \sin\Theta \cos\Omega \\ \sin\Theta\sin\Omega \\ \cos\Theta \end{pmatrix}.
\end{equation}
As the wave function $\psi$ evolves according to \eqref{eq:DirEl}, the unit vector $\vec{n}$ in \eqref{def:n} traces out a path on the unit sphere (often called {\em the Bloch sphere}), providing a convenient way of visualizing the directional changes in $\psi$.

Observe that the third component of the vector in \eqref{def:n} is precisely the Bohmian velocity $v^{(\psi)}$ in \eqref{def:vel}. In addition to the velocity, one can also define local expressions for the momentum $p^{(\psi)}$, and energy $E^{(\psi)}$ of the state $\psi$ in terms of the Cayley-Klein parameters $\Theta$ and $\Omega$ (see Vigier \cite{Vig52}, who attributes this definition to Bohm):

\begin{equation}\label{def:vpE}
v^{(\psi)} = \cos\Theta,\qquad p^{(\psi)} = m\cot{\Theta}\sec\Omega, \qquad E^{(\psi)} = m\csc{\Theta}\sec\Omega.\end{equation}

Note that these are fields defined on the 1-particle configuration space. Thus, similar to the velocity, we can define the momentum and energy of a Dirac particle with trajectory $q(t)$ to be $p^{(\psi)}(t,q(t))$ and $E^{(\psi)}(t,q(t))$, respectively. At this point, however, it is not clear if these will have any connection to the physical quantities of the same names that could be observed in an experiment.  The aim of this paper is to establish such a connection.

\subsection{Classical and Quantum notions of Energy and Momentum}

In classical mechanics, momentum and energy are of great importance in analyzing particle dynamics. We define non-relativistic momentum and kinetic energy as $mv$ and $\frac{1}{2}mv^2$, and in relativistic mechanics \edit{(with $c=1$)}, we adjust these to:

\begin{equation}\label{eq:BohmPandE}p = \gamma m v, \qquad E = \gamma m, \qquad \gamma := \frac{1}{\sqrt{1-v^2}}.
\end{equation}
We can see that these coincide with the quantities we called energy and momentum in \eqref{def:vpE} only in the case $\Omega = 0$, but not otherwise.  This in hindsight should not be surprising since \edit{in quantum mechanics }the law of motion for particles is not the classical one.  In particular the velocity and momentum of a relativistic particle do not have to be aligned, and its energy can be negative.  
From now on, we will refer to the values that the quantities $p^{(\psi)}$ and $E^{(\psi)}$ in \eqref{def:vpE} take on the path of the particle as the \textit{Bohmian momentum} and \textit{Bohmian energy} of that particle.
 
In orthodox quantum mechanics, we replace observable quantities such as $p$ and $E$ with operators on the Hilbert space of wave functions. The momentum operator (in 1-D) is defined as
\begin{equation}
\label{eq:POperator}
P := -i\hbar\frac{\partial}{\partial s}.
\end{equation} 
Consequently, we may try to associate the momentum of the system at some time with the expected value of this operator, which we define using the Hilbert space inner product: 
\begin{equation}\label{eq:ExpValP}
\langle \psi, P \psi \rangle = \int_{\mathbb{R}} \psi^{\dagger} P \psi \,ds.
\end{equation} 
Similarly we may associate the Hamiltonian operator, denoted by $H$, with energy. The Hamiltonian of a system is the operator appearing on the right-hand side of the general Schrodinger Equation that the wave function $\psi$ is supposed to satisfy:
\begin{equation}\label{Schro}
i\hbar\frac{\partial}{\partial t}\psi = H\psi.
\end{equation} We can take its expected value, $\langle \psi, H \psi \rangle$, to find a candidate for the value for the energy of the system as in \eqref{eq:ExpValP}, just as we did for momentum. It is clear however that these must be very different quantities from their Bohmian counterparts.  In particular, since the operators $P$ and $H$ both commute with $H$, both of these expectations are conserved, i.e. independent of time, while the momentum and energy of the particle changes since the guiding velocity $v^{(\psi)}$ is time-dependent in general, as we already observed in the non-relativistic case.  We can still hope that, just as in the non-relativistic case, these Bohmian counterparts have constant {\em asymptotic} values that are the same, or at least close to, the expected values of momentum and energy operators.
\subsection{Spinor Wave Packets}
When solving for trajectories of the electron, we are going to assume that the initial wave function is a {\em Gaussian wave packet}. This means that the two components of the initial wave function are of the form 
\begin{equation} \label{Gaussiandata}
 \mathring{\psi}(s) = \begin{pmatrix}
     \cos\frac{\Theta_0}{2}e^{i\Omega_0/2}\\[5pt]
     \sin\frac{\Theta_0}{2}e^{-i\Omega_0/2}
 \end{pmatrix} \frac{1}{\sqrt[4]{2\pi \sigma^2}} e^{-\frac{s^2}{4\sigma^2} + ik_0s}, 
 \end{equation}
 with $\Theta_0\in[0,\pi]$, $\Omega_0\in[0,2\pi)$, $\sigma>0$, and $k_0\in \mathbb{R}$ free parameters representing the initial Bloch sphere angles, standard deviation of position, and (as we shall see) mean momentum, respectively.
Unlike plane waves, these Gaussian packets are elements of the Hilbert space. 
Of course, they are an idealization, but when plugged into \eqref{eq:DiracCurrent}, they yield a normal  distribution with mean $0$ and standard deviation $\sigma$, which is physically compelling since we know that normal distributions saturate the Uncertainty Principle (UP), so this type of data can represent a particle that, {\em to the extent allowed by the UP}, is localized in both position and momentum.

The main drawback of Gaussian wave packet initial data for Dirac's equation is that, unlike the non-relativistic case of free Schr\"odinger flow, the free Dirac flow (in the massive case) does {\em not} preserve the Gaussian profile. As we shall see, a single wave packet, as it evolves under the Dirac flow, will break up into several ones that will continue to interfere with each other so long as their supports overlap.  The asymptotic analysis of the wave function and its effect on particle trajectories is thus more delicate than what we saw above in the non-relativistic case.

\subsection{Expected Values \edit{of Momentum and Energy} for Wave Packets}
Given a self-adjoint $H$, the general Schrodinger equation in \eqref{Schro} yields a \textit{unitary flow}:

\[\psi(t,s) = e^{-itH/\hbar}\psi(0,s).\]

It follows from the above and the self-adjointness of the Dirac Hamiltonian that the expected value of any self-adjoint operator acting on the wave function is conserved for all time $t$, so long as it commutes with $H$.  This means we can construct Gaussian initial data with a particular expected value for the momentum operator, and we know that this expected value is the same at all later times.

From now on, we will be setting $\edit{c =}\ \hbar = 1$. 

We recall that $P e^{iks} = k e^{iks}$, thus the plane wave $e^{iks}$ is a (pseudo-)eigen function of the operator $P$.  We can likewise find the eigenvalues and eigenfunctions of the Hamiltonian $H$ (the energy operator in QM), and we find $H \psi = E \psi$ for $\psi = \Phi^\pm_k$, i.e. the energy eigenfunctions are parametrized by a real number $k$ and a sign, and that we have
$$ H \Phi^\pm_k = \pm \sqrt{k^2 + m^2} \Phi^\pm_k.$$
In other words, $E = \pm \sqrt{k^2 + m^2}$, i.e. the spectrum of $H$ is the complement of the interval $(-m,m)$ in $\mathbb{R}$.  The eigenfunctions of $H$ are also plane waves:
\begin{equation}\label{def:Phik}
    \Phi^\pm_k(s) := e^{iks} \begin{pmatrix}
        \cos\frac{\Theta_{\pm,k}}{2} e^{i\Omega_\pm/2}\\[5pt]
        \sin\frac{\Theta_{\pm,k}}{2} e^{-i\Omega_\pm/2}
    \end{pmatrix},
\end{equation}
where 
\begin{equation}\label{def:Thetak}
    \Theta_{+,k} := \tan^{-1} \frac{m}{k},\quad \Theta_{-,k} := \pi - \Theta_{+,k}, \qquad
    \Omega_+ := 0,\qquad \Omega_- := \pi.
\end{equation}
Therefore the pair $(\Theta_{\pm,k},\Omega_\pm)$ represent antipodal points on the Bloch sphere.
\begin{remark}
It is important to note that if we only fix the energy $E$, then there will be {\em two} values of $k$ associated with it: $k = \pm \sqrt{E^2 - m^2}$, and the corresponding wave packets to these two will propagate in opposite directions to each other.  In what follows we restrict ourselves to just one of them\edit{.}
\end{remark}
For the Gaussian wave packet \eqref{Gaussiandata}, we can compute the expected value of $P$ as follows:

\begin{eqnarray}\langle \mathring{\psi}(s), P \mathring{\psi}(s) \rangle  & = & \int_{-\infty}^{\infty} \begin{pmatrix}\mathring{\psi}_-^* & \mathring{\psi}_+^* \end{pmatrix} (k_0 + i\frac{s}{2 \sigma^2}) \begin{pmatrix}\mathring{\psi}_- \\ \mathring{\psi}_+ \end{pmatrix} \,ds \\
& = & \frac{1}{\sqrt{2\pi\sigma^2}} \int_{-\infty}^{\infty} k_0 e^{\frac{-s^2}{2\sigma^2}} + i\frac{s}{2 \sigma^2}e^{\frac{-s^2}{2\sigma^2}} \,ds =  k_0.
\end{eqnarray}
Thus for a Gaussian wave packet initial data, the expected value of the momentum of the wave function is $k_0$ at all times which, together with what we already observed in (\ref{initSchrod}, \ref{initSchrodmom}) justifies why we called the parameter $k_0$ {\em mean momentum}.

Similarly, we can compute the expected value of the free Dirac Hamiltonian of the electron with respect to a Gaussian wave packet \eqref{Gaussiandata}: 
\begin{eqnarray}
\langle \mathring{\psi}(s), H_{\text{\text{el}}} \mathring{\psi}(s) \rangle & = & \int_{-\infty}^{\infty} \mathring{\psi}^\dagger \begin{pmatrix} P & m \\ m & -P \end{pmatrix} \mathring{\psi} \, ds \\
& = & \frac{1}{\sqrt{2\pi\sigma^2}}\int_{-\infty}^{\infty}e^{\frac{-s^2}{2\sigma^2}}\big{(}\cos{\Theta_0}(k_0+i\frac{s}{2\sigma^2}) + \frac{1}{2}m\sin{\Theta_0}(e^{i\Omega_0}+e^{-i\Omega_0})\big{)} \,ds \nonumber\\
  & = & k_0 \cos{\Theta_0} + m \sin{\Theta_0}\cos{\Omega_0}\ \edit{\in [-\sqrt{k_0^2+m^2},\sqrt{k_0^2+m^2}].} 
  \end{eqnarray}
Thus, if we base our Gaussian wave packet on the energy eigenfunctions, i.e. let $\Theta_0 = \Theta_{\pm,k}$ and $\Omega_0 = \Omega_\pm$, so that
\begin{equation}
    \mathring{\Psi}^\pm_E(s) := \Phi^\pm_k \frac{1}{\sqrt[4]{2\pi \sigma^2}} e^{-\frac{s^2}{4\sigma^2}},\qquad k^2 + m^2 = E^2,\qquad E\geq m > 0,
\end{equation}
then we have 
$$
\langle \mathring{\Psi}_E^\pm, H \mathring{\Psi}_E^\pm \rangle = \pm E,
$$
and thus since energy is conserved under the unitary flow generated by $H$, we have that the expected value of the energy of the solution $e^{itH}\Psi_E^\pm$ at all later times is still $\pm E$.  

Given $E\geq m>0$ let us fix one of the two possible values of $k = \pm\sqrt{E^2 - m^2}$ \edit{so that} 
\begin{eqnarray}
 \label{posen}   \mathring{\Psi}^{+}_E & = & \frac{1}{\sqrt[4]{2\pi \sigma^2}}e^{-\frac{s^2}{4\sigma^2}+  i k s}\begin{pmatrix}
       \edit{\sqrt{(E+k)/2E}}\\ \edit{\sqrt{(E-k)/2E}}
     \end{pmatrix}\\
 \label{negen}   \mathring{\Psi}^{-}_E & = &\frac{i}{\sqrt[4]{2\pi \sigma^2}}e^{-\frac{s^2}{4\sigma^2} +  i k s}\begin{pmatrix}
        \edit{\sqrt{(E-k)/2E}}\\ - \edit{\sqrt{(E+k)/2E}}
    \end{pmatrix}.
\end{eqnarray}
We shall see that even if we start with only one of these, say the positive energy one $\Psi^+_E$, the Dirac dynamics give birth to a wave packet with negative energy $-E$, which will however transport in the opposite direction of the positive energy part, since the two packets will have the same momentum $k$, and the velocity of negative energy packets is in the opposite direction of their momentum. We 
\edit{show} that in that case, after sufficient time has passed, the supports of these two packets will have negligible overlap, causing a particle with a {\em typical} initial position to be \edit{in effect} guided by only one of the two, depending on whose support the particle finds itself in at any given time.  

If on the other hand we take initial data with a fixed expected value of energy $E$ that employs both possible values of $k$, it could yield a highly oscillatory trajectory for the electron, the reason being that the part of the wave with energy $E$ and momentum $k$ and the part with energy $-E$ and momentum $-k$, will not separate but travel together, so that they will continue to interfere with each other\footnote{We thank Roderich Tumulka for pointing this out to us.}. We shall not be considering this case here.

We can now state the main result of this paper:
\begin{theorem} \label{thm:main}
Given initial data of the form 
\begin{equation}\label{init}
    \mathring{\psi}(s) = \frac{1}{\sqrt[4]{2\pi \sigma^2}}e^{-\frac{s^2}{4\sigma^2} +  i k s}\begin{pmatrix}
         \edit{a_-}\\ \edit{a_+} \end{pmatrix},\qquad \edit{a_\pm \in \mathbb{C},\quad |a_-|^2+|a_+|^2 = 1,}
\end{equation}
with \edit{expected value of} momentum $k \in \mathbb{R}$, $k \ne 0$, the solution to the initial value problem for the Dirac equation (\ref{eq:DirEl}, \ref{initdat}) as $t\to \infty$ asymptotically approaches the superposition of two wave packets with \edit{average} energies \edit{$\sqrt{m^2+k^2}$ and $-\sqrt{m^2+k^2}$}, both with \edit{average} momentum $k$, that travel in opposite directions. \edit{Moreover}, the \edit{trajectory} of a typical particle guided by this wave \edit{is} eventually \edit{close to} a straight line, with the particle traveling with \edit{an approximately} constant velocity to the left or to the right, depending on its initial position. 
\end{theorem}
The proof of the above theorem relies on the Stationary Phase Approximation (SPA) method, applied at a fixed, sufficiently large time $t$ and arbitrary position $s$ to the components $\psi_\pm$ of the solution (\ref{psipmsol}) of the Dirac equation.  Under the assumptions of the theorem these are oscillatory integrals, with the frequency of the oscillation proportional to the parameter $\omega = \frac{mc}{\hbar}$.  Thus to apply SPA we need to study the limit $\omega \to \infty$, a regime known as {\em the classical limit} of the theory (see e.g. \cite{RoemerThesis} for definition and references.) \edit{This will be accomplished in the next section.}

\section{Classical Limit of the Dirac Equation}
The free Dirac equation for a single particle in one space dimension, in natural units where $\hbar = \edit{m} = c = 1$,
is
\begin{equation}\label{eq:DirElnat}
-i\gamma^{\mu} \frac{\partial}{\partial x^{\mu}}\psi + \psi = 0.  
\end{equation}
In Hamiltonian form, this becomes
\begin{equation}\label{1dDir}
    i \partial_t \psi = - i \alpha \partial _s \psi + \beta \psi,
\end{equation}
where $\alpha := \gamma^0 \gamma^1$ and $\beta = \gamma^0$.  
We assume that initial data has been prescribed for this at $t=0$:
$$
\psi(0,s) = \mathring{\psi}(s).$$
In the units we have picked, the Compton Wavelength of the electron, $\lambda_{CWE} = 1$.  To study the classical limit, we imagine that there is a characteristic macroscopic length $L \gg 1$ associated with the system (e.g. distance of a detector from the source for the particle.)  We introduce {\em macroscopic variables} 
$$ t' = \frac{t}{L},\qquad s' = \frac{s}{L},$$
and we define the wave function in these variables as
$$ \psi^L(t',s') := \sqrt{L} \psi(Lt',Ls').$$
This definition assures that for normalized wavefunctions $\psi$, we have
$$ \int_{-\infty}^\infty |\psi^L(t',s')|^2 ds' = \int_{-\infty}^\infty |\psi(Lt',s)|^2 ds = 1 \qquad \forall t'\in \mathbb{R},$$
so that normalization is preserved. 
The equation satisfied by $\psi^L$ is
\begin{equation}\label{eq:psiL}
    i \partial_{t'} \psi^L = -i \alpha \partial_{s'} \psi^L + \omega {\beta} \psi^L,
\end{equation}
where 
$$ \omega := \frac{mc}{\hbar} L = \frac{L}{\lambda_{CWE}}\gg 1.$$
Thus $\omega$ is a large dimensionless parameter. 
The corresponding initial data is
$$\psi^L(0,s') = \mathring{\psi^L}(s') = \sqrt{L} \mathring{\psi}(L s').
$$
We would like to study the limit $\omega \to \infty$.  How does the wave function look like in this limit? And what about the trajectories?  

For the latter question, we need the guiding equation
$$
\frac{dQ}{dt} = v^\psi(t,Q(t)) = \frac{|\psi_-|^2 - |\psi_+|^2}{|\psi_-|^2 + |\psi_+|^2}(t,Q(t)),\qquad Q(0) = q_0.
$$
In macroscopic variables, this is virtually unchanged:
$$
\frac{dQ^L}{dt'} = v^{\psi^L}(t',Q^L(t')) =  \frac{|\psi^L_-|^2 - |\psi^L_+|^2}{|\psi^L_-|^2 + |\psi^L_+|^2}(t',Q^L(t')),\qquad Q^L(0) = q'_0 = q_0/L
$$
\subsection{Solution of the Dirac equation in macroscopic variables}
\edit{By \eqref{psipmsol}} the solution to \eqref{eq:psiL} is (dropping the primes on the variables from now on)
\begin{align}
    \psi^L_{\pm}(t,s) &= \overset{\circ}{\psi^L_{\pm}}(s \pm t) \nonumber\\ &- \frac{\omega}{2}\int_{s-t}^{s+t} J_1\left(\omega\sqrt{t^2 - (s - \sigma)^2}\right)\frac{\sqrt{t \mp (s- \sigma)}}{\sqrt{t \pm (s- \sigma)}}\overset{\circ}{\psi^L_{\pm}}(\sigma) \,d\sigma\nonumber \\ &- \frac{i\omega}{2} \int_{s-t}^{s+t} J_0\left(\omega\sqrt{t^2 - (s - \sigma)^2}\right) \overset{\circ}{\psi^L_{\mp}}(\sigma) \,d\sigma.
\end{align}
We assume that the initial data for \eqref{eq:psiL} is a Gaussian wave packet \eqref{posen} with energy $$E = \omega E_0> \omega$$ and momentum $$k = \omega p_0 \ne 0.$$  (Note that $m=\omega$ now.)

\edit{The initial data \eqref{init} is of the form}
\begin{equation}\label{initmixed}
    \mathring{\Psi^L} = a_- \mathring{\Psi}^{L}_- + a_+ \mathring{\Psi}^{L}_+, \qquad |a_-|^2+|a_+|^2 = 1,
\end{equation}
where
\begin{equation}\label{initial}
\mathring{\Psi}^L_-(s) = \begin{pmatrix}
1\\
0
\end{pmatrix}
e^{i\omega p_0 s}f_{\sigma}(s),
\qquad
\mathring{\Psi}^L_+(s) = \begin{pmatrix}
0\\
1
\end{pmatrix}
e^{i\omega p_0 s}f_{\sigma}(s),
\end{equation}
and the amplitude $f_\sigma :\mathbb{R} \to \mathbb{R}$ is such that its square is the density function of a normal distribution with mean $0$ and standard deviation $\sigma$:
\begin{equation}\label{cond:f}
f_\sigma(\edit{s}) := \frac{1}{\sqrt[4]{2\pi\sigma^2}} \exp{\frac{-\edit{s}^2}{4\sigma^2}}.
\end{equation}
For this to be a faithful model of the dynamics of a particle, we may also assume 
\begin{equation}\label{mainass}
    \frac{1}{\omega p_0} \ll \sigma \ll 1,
\end{equation}
which states that the wavelength of the particle is much smaller than the width of the Gaussian, which in turn is much smaller than the macroscopic length scale.

By the linearity of Dirac's equation, and the symmetry of the problem under parity, it is enough to \edit{analyze} the solution corresponding to just one of \edit{$\mathring{\Psi}^L_\pm$, since for example} the solution for \edit{$\mathring{\Psi}^L_-$ data} can be obtained from the one \edit{for $\mathring{\Psi}^L_+$ data}.

\edit{The first component} of $\mathring{\Psi}^L_+$ \edit{being zero}, the components $\psi_{\pm}$ of the \edit{corresponding} solution are: 
\begin{equation}\label{sol:psiplus}
\psi_{+}(t, s) = e^{ip_0\omega(s+t)}f(s+t) - \frac{\omega}{2} \int_{s-t}^{s+t} \frac{J_1(\omega\sqrt{t^2 - (s-\sigma)^2)}}{\sqrt{t^2 - (s- \sigma)^2}}(t - (s- \sigma))f(\sigma)e^{ip_0\omega\sigma} \,d\sigma,
\end{equation}
and
\begin{equation}\label{sol:psiminus}
\psi_-(t, s) = -\frac{i\omega}{2}\int_{s-t}^{s+t} J_0(\omega\sqrt{t^2 - (s-\sigma)^2})f(\sigma)e^{ip_0\omega\sigma} \,d\sigma.
\end{equation}
(For the sake of brevity we have dropped the superscript $L$ and subscript $\sigma$ on these formulas.)  We see that $\psi_+$ is a sum of a transport term and an integral term.  It turns out that there is a cancellation between these two terms, which can be used to eliminate all or part of the transport term.  In order to see this, we introduce the following lemma\footnote{We thank Lawrence Frolov for giving us this idea.}: 
\begin{lemma} \label{lem:trans}
Let $j_0 \approx 2.404825$ denote the smallest positive zero of the Bessel function $J_0$. 
For all $\alpha \in (0,1]$ there exists $\beta = \beta(\alpha)\in (0,1]$ such that if we define
\begin{equation}
    \tilde{\psi}_+(t,s) :=(1-\alpha) e^{ip_0\omega(t+s)}f(t+s) -\frac{\omega}{2}\int_{s-t}^{s+t - 2t\beta}\frac{J_1(\omega\sqrt{t^2 - (s-\sigma)^2})}{\sqrt{t^2 - (s- \sigma)^2}}(t - (s- \sigma))f(\sigma)e^{ip_0\omega\sigma} \,d\sigma,
\end{equation}
then there exists a constant $C>0$ such that for all $s \in \mathbb{R}$ and every $\omega,t>0$ such that $\omega t > j_0/2$,  we have
$$\left\vert\psi_+(t,s) - \tilde{\psi}_+(t,s)\right\vert < \frac{C|p_0|/\sqrt{\sigma}}{\omega t}.$$
Moreover, $\beta(\alpha) \in (0, \frac{j_0^2}{4 (\omega t)^2})$, and $\beta$ can be chosen to be a continuous non-decreasing function of $\alpha$.
\end{lemma}
\begin{proof} 
Let $\chi := \psi_+ - \tilde{\psi}_+$.  Then
$$
\chi = (\alpha - A(\beta)) \phi(t+s) -\frac{\omega}{2}\int_{s+t - 2 \beta t}^{s+t} \frac{J_1(\omega\sqrt{t^2 - (s-\sigma)^2})}{\sqrt{t^2 - (s- \sigma)^2}}(t - (s- \sigma))\left( \phi(\sigma)-\phi(s+t) \right)\,d\sigma,
$$
where $\phi(x) := f(x)e^{ip_0\omega x}$, with $f$ as in \eqref{cond:f}, and
$$
A(\beta) := \frac{\omega}{2} \int_0^{2 \beta t}
 \frac{J_1(\omega\sqrt{2t\tau - \tau^2})}{\sqrt{2t\tau - \tau^2}}(2t - \tau)  \,d\tau.
$$ 
Using the assumption on the initial data profile $f$:
$$|\phi(s + t - \tau) - \phi(s+t)| \leq \sup_\mathbb{R}|\phi'|\ \tau  \leq (\frac{1}{\sigma^{3/2}}+\frac{\omega |p_0|}{\sigma^{1/2}}) \tau =: M\tau.$$ 
Thus we have $\chi = (\alpha - A) \phi(s+t) + B$ where
 $$ B = - \frac{\omega}{2} \int_0^{2\beta t} \frac{J_1(\omega\sqrt{2t\tau - \tau^2})}{\sqrt{2t\tau - \tau^2}}(2t - \tau) (\phi(s+t - \tau) - \phi(s+t)) \,d\tau,$$
 so that using the simple estimate $|J_1(\omega x)/x| \leq \omega/2$ for all $x>0$ we obtain
 \begin{eqnarray*}
     |B| & \leq & 2tM(\omega t)^2 \int_0^{\beta} z(1-z) dz \leq t M (\omega t)^2 \beta^2 \leq \frac{C M t}{(\omega t)^2}, 
 \end{eqnarray*}
 (with $C$ henceforth denoting a generic positive constant) provided, as we shall see below, $\beta < C/(\omega t)^2$.  Recalling the definition of $M$, it then follows that $$|B|< C|p_0|/(\sqrt{\sigma}\omega t).$$
 On the other hand, by the Taylor series for $J_1$ we have that
 \begin{eqnarray*}
  A   & = & \sum_{m=0}^\infty \frac{ (-1)^m (\omega t)^{2m+2} }{m! (m+1)!} \int_0^{\beta} (1-z)^{m+1} z^m dz \\
     & = & \sum_{m=0}^\infty \frac{ (-1)^m (\beta (\omega t)^2)^{m+1} }{(m+1)! (m+1)!} - \sum_{m=0}^\infty \frac{ (-1)^m (\omega t)^{2m+2} }{m! (m+1)!} \int_0^{\beta} \frac{1 - (1-z)^{m+1}}{z} z^{m+1} dz,
 \end{eqnarray*}
and by the Taylor series expansion for $J_0$ we have
 $$ \sum_{m=0}^\infty \frac{ (-1)^m (\beta (\omega t)^2)^{m+1} }{(m+1)! (m+1)!} = 1- \sum_{m=0}^\infty \frac{(-1)^m (\beta (\omega t)^2)^m}{(m!)^2} = 1 - J_0(2\omega t \sqrt{\beta}) =: g(\beta).$$
 The function $g$ is clearly continuous, monotone increasing, and $g(0) = 0$, $g(j_0^2/4 (\omega t)^2) = 1$.  Thus there exists a value of $\beta \in (0, \frac{j_0^2}{4 (\omega t)^2})$, call it $\beta(\alpha)$, such that $g(\beta(\alpha)) = \alpha$ for any $\alpha \in [0,1]$.
 Meanwhile,
 \begin{eqnarray*}
    0 \leq \int_0^\beta\frac{1 - (1-z)^{m+1}}{z} z^{m+1} dz 
     & \leq & \int_0^\beta \sup_{0\leq z \leq \beta} \left|\frac{1 - (1-z)^{m+1}}{z} \right| z^{m+1} dz \\
     & \leq & \frac{m+1}{m+2}\beta^{m+2}.
 \end{eqnarray*}
 Thus
 assuming $\omega t > j_0/2$ we obtain, since $\beta(\alpha) < j_0^2/4(\omega t)^2$,
 $$ |A - \alpha| \leq  \frac{1}{(\omega t)^2}\sum_{m=0}^\infty \frac{(\beta(\alpha) (\omega t)^2)^{m+2}}{m! m! (m+2)} \leq \frac{D}{(\omega t)^2}, \qquad D \approx 2.36353,$$
which establishes the claim.  
\end{proof}
\subsection{The Stationary Phase Approximation}
Using the two-dimensional version of the stationary phase approximation method (see e.g. Chako \cite{Chako}) we can  find approximations for $\psi_-$ and $\psi_+$. 

\begin{theorem}\label{thm:SPA}
For all $p_0\ne 0$, $T>0$ and $\sigma>0$, there exists $\Omega = \Omega(p_0,T,\sigma)>0 $ such that if $\psi_\pm$ given by \eqref{sol:psiminus} and \eqref{sol:psiplus} are the components of the wave function $\psi$ of a single electron corresponding to initial data of the form 
\begin{equation}
    \mathring{\psi}(s) = \frac{1}{\sqrt[4]{2\pi\sigma^2}}e^{- \frac{s^2}{4\sigma^2} + i\omega p_0 s} \begin{pmatrix}
        0\\ 1
    \end{pmatrix},
\end{equation}
then for all $\omega>\Omega$, all $t \in [\frac{|v_0|}{2}T, T]$ 
and  all $s\in \mathbb{R}$ 
we have
\begin{equation}\label{UandR}
  \psi_\pm(t,s) = U_1^\pm(t,s)  + R_{1,1}^\pm(t,s),  
\end{equation}
where
\begin{equation}\label{Uminus}
    U_1^-(t,s) :=
    \frac{1}{2E_0}\left[f(s-v_0t)e^{i\omega (p_0 s - E_0t)} - f(s+v_0t)e^{i\omega (p_0 s + E_0t)}\right]
\end{equation}
and
 \begin{equation}\label{Uplus}
    U_1^+(t,s) := 
    \frac{1}{2E_0}\left[f(s-v_0t)(E_0-p_0)e^{i\omega (p_0 s - E_0t)} + f(s+v_0t)(E_0+p_0)e^{i\omega (p_0 s + E_0t)} \right]
\end{equation}
with 
$$ 
E_0 := \sqrt{1+p_0^2},\qquad v_0 := \frac{p_0}{\sqrt{1+p_0^2}}
$$
are the Stationary Phase Approximations (SPA) to $\psi_-$ and $\psi_+$ respectively.
Moreover, there are numerical constants 
$A,B>0$ such that
\begin{equation}\label{errorest}
|R_{1,1}^\pm(t,s)| \leq A\frac{|p_0|^5 e^{Bt/\sigma}}{\sqrt{\omega}}.
\end{equation}
Thus for any $T>0$ the error in the SPA can be made arbitrarily small for all $t \in [\frac{|v_0|}{2}T,T]$ provided $\omega$ is sufficiently large.
\end{theorem}
The proof of this Theorem is given in the Appendix. 
Using the left-right symmetry inherent in the problem, we then obtain
\begin{corollary}\label{cor:mixed}
    In the more general case where the initial data is as in \eqref{initmixed}, 
     the stationary phase approximation to $\psi_\pm$ is
    \begin{equation}
        U_\pm = \edit{a_+} U_1^\pm + \edit{a_-} U_2^\pm,
    \end{equation}
    where $U_1^\pm$ are as in \eqref{Uminus}, \eqref{Uplus}, 
    and $U_2^\pm$ can be obtained  by replacing $$ s \to -s,\qquad v_0 \to -v_0,\qquad p_0 \to -p_0,$$ in the formulas for $U_1^\mp$ and taking into account that $f(s) = f(-s)$, i.e.
 \begin{equation}\label{Uminus2}
    U_2^-(t,s) := 
    \frac{1}{2E_0}\left[f(s+v_0t)(E_0-p_0)e^{i\omega (p_0 s + E_0t)} + f(s-v_0t)(E_0+p_0)e^{i\omega (p_0 s - E_0t)} \right],
\end{equation}
and    \begin{equation}\label{Uplus2}
    U_2^+(t,s) :=
    \frac{1}{2E_0}\left[-f(s+v_0t)e^{i\omega (p_0 s + E_0t)} + f(s-v_0t)e^{i\omega (p_0 s - E_0t)}\right].
\end{equation}
Thus 
the full SPA for $\psi$ is as follows:
\begin{equation}\label{Ufull}
     U =  \frac{1}{2E_0} \begin{pmatrix}
      \edit{a_+}  +   \edit{a_-} (E_0+p_0) \\
      \edit{a_-}  +   \edit{a_+} (E_0-p_0)
      \end{pmatrix}      \phi_- +
      \frac{1}{2E_0} \begin{pmatrix}
      -\edit{a_+}  +   \edit{a_-} (E_0-p_0)\\
      -\edit{a_-}  +   \edit{a_+} (E_0+p_0)
     \end{pmatrix} \phi_+,
\end{equation}
where 
\begin{equation}\label{def:phipm}
    \phi_+ := f(s+v_0t)e^{i\omega (p_0 s + E_0t)}, \qquad \phi_- := f(s - v_0t)e^{i\omega (p_0 s - E_0t)}.
\end{equation}
Here $\phi_-$ is the wave packet with \edit{average} momentum $\omega p_0$ and \edit{average} energy $\omega E_0$ that is traveling with velocity $v_0$ in the same direction as the direction of the momentum, while $\phi_+$ has \edit{average} momentum $\omega p_0$ and energy $-\omega E_0$, and is traveling with velocity $-v_0$, i.e. in the opposite direction of the momentum. 
\end{corollary}
\vspace{-10pt}
\edit{
\begin{remark}
    From \eqref{Ufull} it is easy to check that for initial data corresponding to the positive energy wave packet $\mathring{\Psi}^+_E$ defined in \eqref{posen} the stationary phase approximation contains only the $\phi_-$ term, and similarly for initial data $\mathring{\Psi}_E^-$ as in \eqref{negen} only the $\phi_+$ term survives.
\end{remark}
}
We thus note that, so long as $p_0$ is non-zero, with the passage of time the centers of the two packets $\phi_-$ and $\phi_+$ will be \edit{so} far from each other that the overlap of their supports would be negligible, which allows us to establish the claims of Theorem \ref{thm:main} about the trajectories.  We will do so in the next section.
\subsection{Analysis of Trajectories}
We turn our attention to the guiding equation \eqref{eq:guiding}, taking its right-hand-side to be the velocity field
corresponding to the stationary phase approximation $U$ of the wave function $\psi$ satisfying \eqref{eq:psiL}, with initial data of the form \eqref{initmixed}.  For the sake of simplicity, we will only do this for \edit{$a_- = 1, a_+ = 0$. The general case will follow from this by applying a suitable unitary transformation}. By Corollary \ref{cor:mixed} we thus have
\begin{equation}
    U =  \begin{pmatrix} U_- \\ U_+ \end{pmatrix} = \frac{1}{2E_0} \begin{pmatrix}
       (E_0+p_0) \phi_- + (E_0-p_0) \phi_+  \\
      \phi_- - \phi_+
     \end{pmatrix}, 
\end{equation}
where $\phi_\pm$ are defined in \eqref{def:phipm}. Recalling the relations \eqref{def:vpE} with $\Omega_0 = 0$,
\begin{equation}
   v_0 = \cos\Theta_0,\qquad p_0 = \cot\Theta_0,\qquad E_0 = \csc\Theta_0,
\end{equation}
we have
\begin{equation}\label{UpmSPA}
    U_- = \cos^2\frac{\Theta_0}{2} \phi_- + \sin^2\frac{\Theta_0}{2}\phi_+,\qquad U_+ = \sin\frac{\Theta_0}{2}\cos\frac{\Theta_0}{2}(\phi_- - \phi_+).
\end{equation}
The corresponding velocity field is therefore
\begin{equation}\label{velU}
    v^{(U)}(t,s) = \cos\Theta_0 \frac{\cos^2\frac{\Theta_0}{2} |\phi_-|^2 - \sin^2\frac{\Theta_0}{2}|\phi_+|^2}{\cos^2\frac{\Theta_0}{2} |\phi_-|^2 + \sin^2\frac{\Theta_0}{2}|\phi_+|^2}  + \sin\Theta_0 \frac{2 \mbox{Re}(\phi_-^*\phi_+) \sin\frac{\Theta_0}{2}\cos\frac{\Theta_0}{2}}{\cos^2\frac{\Theta_0}{2} |\phi_-|^2 + \sin^2\frac{\Theta_0}{2}|\phi_+|^2}.
\end{equation}
From the above it's already clear that, once the supports of $\phi_-$ and $\phi_+$ are disjoint, if the particle finds itself in the support of $\phi_-$, its velocity will be $v_0$, and if it's in the support of $\phi_+$, the velocity will be $-v_0$.  We now need an ODE argument to turn this into a statement about trajectories.

Consider the guiding equation
\begin{equation}\label{Uode}
    \frac{ds}{dt} = v^{(U)}(t,s),
\end{equation}
where the velocity field $v^{(\psi)}$ on the right hand side of \eqref{eq:guiding} is replaced with $v^{(U)}$ in \eqref{velU}. Without loss of generality, we are going to assume $p_0>0$, \edit{so that $0\leq \Theta_0< \frac{\pi}{2}$}.   We introduce some new variables to simplify this ODE:
\begin{equation}
    x := \frac{\sqrt{v_0}}{\sigma}t,\quad y := \frac{\sqrt{v_0}}{\sigma}s,\quad \eta:= \tan\frac{\Theta_0}{2},\quad  a:= \frac{2\sigma E_0}{\sqrt{v_0}}.
\end{equation}
Using \eqref{cond:f} the ODE then becomes
\begin{equation}\label{xyODE}
    \frac{dy}{dx} = \cos\Theta_0\frac{e^{2xy}-\eta^2}{e^{2xy}+\eta^2}+\sin\Theta_0\frac{2\eta e^{xy}}{e^{2xy}+\eta^2}\cos(a \omega x) := F(x,y).
\end{equation}

We can make the above statement about trajectories quantitative by finding {\em barriers} for solutions of \eqref{xyODE}: We look for curves $y = B_\pm(x)$ in the $xy$ plane with the property that
\begin{equation}\label{Fregions}
    \edit{F(x,B_-(x))\leq  0 \leq B'_-(x)},\qquad\mbox{and}\qquad \edit{F(x,B_+(x))\geq 0 \geq B'_+(x)}.
\end{equation}
If such curves are found, the region in the $xy$ plane with $y \leq B_-(x)$ would be an invariant region, meaning it is mapped into itself under the forward flow, and similarly the region $y\geq B_+(x)$.  Also in these two regions, $y$ becomes a monotone function of $x$, so that in the former case, the trajectory continues towards the left, and in the latter case, towards the right.  For fixed $x>0$, it is evident that as $|y|$ becomes large, the oscillatory term in the right-hand-side $F(x,y)$ in \eqref{xyODE} dies out and we have
\begin{equation}
    \lim_{y\to \infty} F(x,y) = \cos\Theta_0 = v_0,\qquad \lim_{y \to -\infty} F(x,y) = -\cos\Theta_0 = -v_0.
\end{equation}
Hence the velocity, and therefore the momentum, of the particle becomes a constant, as advertised.

To find the $B_\pm$ curves, we first find $y_0(x)$ such that $F(x,y_0(x)) = 0$.  Let $X := e^{xy}>0$. Then
$$
F(x,y) = \frac{\cos\Theta_0}{X^2 +\eta^2} (X^2 + 2\tan\Theta_0 \cos(a\omega x) \eta X - \eta^2)
$$
so that if we define
$$
y_0(x) = \frac{1}{x}\ln\frac{\tan\frac{\Theta_0}{2}}{\tan\Theta_0 \cos a\omega x + \sqrt{1+ \tan^2\Theta_0 \cos^2 a\omega x }}.
$$
we would then have $F(x,y_0(x)) = 0$ and moreover 
$$
y \leq y_0(x) \implies F(x,y)\leq 0,\qquad y\geq y_0(x) \implies F(x,y)\geq 0.
$$
We may thus define \edit{$B_\pm(x) := \frac{C_\pm(\Theta_0)}{x}$ where}
\begin{equation}
   C_- :=\ln\frac{\tan\frac{\Theta_0}{2}}{ \sqrt{1+ \tan^2\Theta_0}+\tan\Theta_0 },
    \qquad \edit{C_+ :=\max\left\{ 0, \ln\frac{\tan\frac{\Theta_0}{2}}{ \sqrt{1+ \tan^2\Theta_0}-\tan\Theta_0 }\right\}},
\end{equation}
so that $B_-(x) \leq y_0(x) \leq B_+(x)$ and \edit{$F(x,B_-(x)) \leq 0 \leq F(x,B_+(x))$}. 

\begin{figure}
    \centering
    \includegraphics[width=0.4\linewidth]{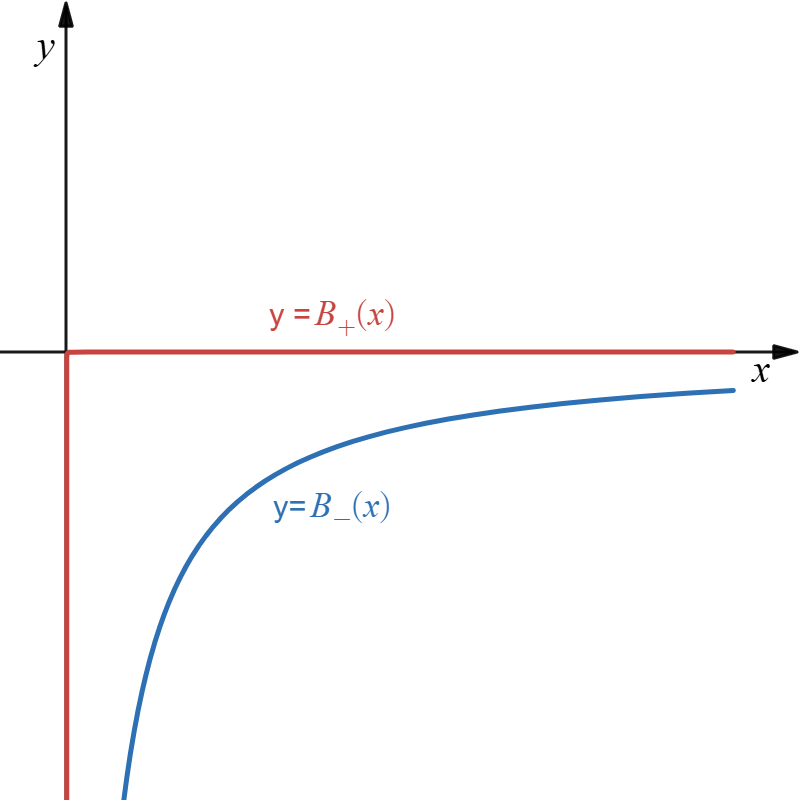}\qquad\includegraphics[width=0.4\linewidth]{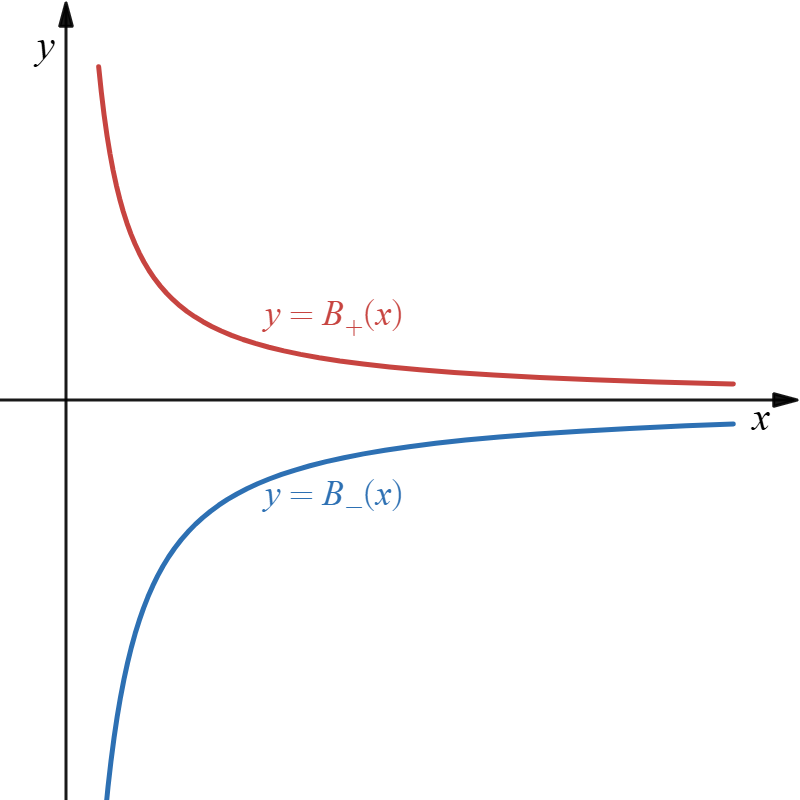}
    \caption{The curves $y = B_\pm(x)$ for $\Theta_0 = \frac{\pi}{8}$ (on the left) and $\Theta_0 = \frac{3\pi}{8}$ (on the right).}
    \label{fig:barriers}
\end{figure}

Next we observe that the curves $y=B_\pm(x)$ are branches of hyperbolas $xy = C_\pm$, \edit{except that for $\Theta_0 \leq \frac{\pi}{4}$}, we have $C_+ = 0$, so that the hyperbola degenerates to $xy=0$. Also note that $C_-<0$ \edit{and $0 \leq C_+\leq |C_-|$} for all $\Theta_0$, 
so that $B_-'(x)>0$ and $B_+'(x) \leq 0$
(see Fig.~\ref{fig:barriers}.) \edit{Therefore \eqref{Fregions} holds.}
In either case, it is clearly true that for large enough $x$, $B_\pm(x)$ can be made arbitrarily small.  We use this, combined with the fact that for large times the supports of $\phi_+$ and $\phi_-$ become almost disjoint, to establish our claim.

\begin{figure}
\input{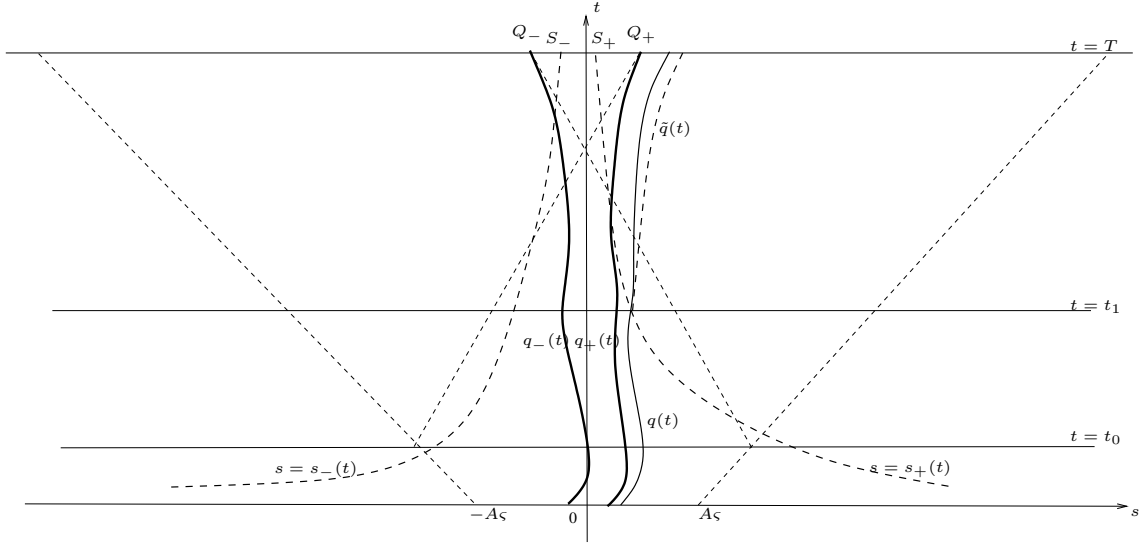}
\caption{\label{fig:Qpm} Barrier construction in the case $\Theta_0>\pi/4$}
\end{figure}

{\bf Proof of Theorem~\ref{thm:main}}:  Let $\epsilon>0$ be given.  In the original variables, the two hyperbolas $y = B_\pm(x)$ are given by $$ s_\pm(t) := \frac{\sigma^2 C_\pm}{v_0 t}.$$
For a given $\epsilon'>0$ let $T_1(\epsilon') = \frac{4\sigma^2 |C_-|}{\epsilon' v_0}$. For fixed $T>0$ let $S_\pm := s_\pm(T)$.  Then for $T\geq T_1$ we have $ -\frac{\epsilon'}{4}<S_-<0$ and $S_-<S_+<-S_-$. Thus $S_+ - S_-<\frac{\epsilon'}{2}$. See Fig.~\ref{fig:Qpm}.

On the other hand, let $A_0>0$ be such that for a random variable $X$ having a normal distribution with mean $\mu = 0$ and standard deviation $\sigma$, i.e. the same as the initial probability distribution $\rho(0,s)$, we have $\mbox{Prob}(|X - \mu|>A_0\sigma) = \frac{\epsilon}{2}$.   Let $t_0 := \frac{v_0}{2}T>0$, and fix an $A>A_0$.  From the hyperbolicity of Dirac's equation, it follows that at time $t_0$, the support of any solution initially supported in the interval $(-A\sigma,A\sigma)$ can at most spread into the interval $(-A\sigma-t_0,A\sigma+t_0)$ (since we have set the speed of light $c=1$).  Since the SPA formula \eqref{UpmSPA} are valid in $[t_0,T]$, $U$ is a linear combination of Gaussian wave packets $\phi_\pm$, which we know travel with speed $v_0$ in opposite directions.   
Let $T>0$ be such that 
$$-v_0 (T - t_0) + A\sigma + t_0 = S_- - \frac{\epsilon'}{4} =: Q_-,\qquad v_0(T - t_0) -A\sigma - t_0 = -Q_-=:Q_+.$$ 
If  $T<T_1$, we simply increase $A$ until $T=T_1$, thus without loss of generality we have $T\geq T_1$.  
As a result the interval $[Q_-,Q_+]$ properly contains $[S_-,S_+]$ and $Q_+ - Q_- \leq \epsilon'$.  
 Now we choose $\epsilon'$ small enough such that 
$$\mbox{Prob}_{\psi_T}(Q_-<s<Q_+) := \int_{Q_-}^{Q_+} |\psi|^2(T,s) ds \leq \frac{\epsilon}{2}.$$

Since the Gaussian packet $\phi_+$ travels to the left with velocity $-v_0$ and the Gaussian packet $\phi_-$ travels to the right with velocity $v_0$, the above ensures that at time $T$ the supports of $\phi_\pm$ are almost disjoint, in the sense that the probability of being in the overlap region is less than $\epsilon/2$.  

Consider now two trajectories (guided by the original wave function $\psi$) $s = q_\pm(t)$ for $t\in [0,T]$ with {\em final} data $q_\pm(T) = Q_\pm$. These trajectories exist, are unique, and at time $t=0$ will have initial values $q_\pm(0)$ in the interval $(-A\sigma,A\sigma)$.
From the equivariance property of $\rho(t,s) = |\psi(t,s)|^2$ it follows that 
$$
\int_{q_-(0)}^{q_+(0)} \rho(0,s) ds = \int_{Q_-}^{Q_+} \rho(T,s) ds \leq \frac{\epsilon}{2}.
$$
We now recall the notion of {\em typicality} (see D\"urr et. al. \cite{DGZ92}): 
\begin{definition}
 A statement about a system of particles governed by the wave function $\Psi_t(s)$ is true for {\em typical} initial configurations if the set of initial configurations for which it is false is small in the sense provided by the {\em quantum equilibrium distribution} $\mathbf{P}^{\Psi_0}(ds) := |\Psi_0(s)|^2 ds$.   
\end{definition}
Let $\mathcal{A}_\epsilon:= (-A\sigma,q_-(0))\cup (q_+(0),A\sigma)$.  We thus have that for any given $\epsilon>0$, $\mbox{Prob}(X \notin \mathcal{A}_\epsilon) < \epsilon$.  Our typicality claim in Theorem~\ref{thm:main} would thus be established if we manage to prove the statements of the theorem for all trajectories whose initial data belong to $\mathcal{A}_\epsilon$.

By uniqueness of solutions to ODEs, any trajectory $q(t)$ that starts to the right of $q_+(0)$ must remain to the right of $q_+(t)$ for all $t$, 
and thus at time $t=T$, it is to the right of $Q_+$. At time $t_0:=\frac{v_0}{2}T$ this trajectory enters the region of spacetime where the stationary phase approximation is valid.  Let $t_1 \in [t_0,T)$ denote the largest time at which $q(t)$ crosses the hyperbola $\mathcal{H}_+ := \{(t,s)\ |\ s = s_+(t)\}$ from left to right.  If no such crossing exists we set $t_1 = t_0$.   Consider then the trajectory $\tilde{q}(t)$ that is guided by $U$ instead of $\psi$, starting at $t=t_1$ with initial value $\tilde{q}(t_1) = q(t_1)$.  Since the hyperbola $\mathcal{H}_+$ is a barrier for solutions of \eqref{Uode}, $\tilde{q}$ cannot cross it back, and is forced to move monotonically to the right. It will thus be (almost) entirely guided by $\phi_-$ and will have \edit{an approximately} constant velocity $v_0>0$ (recall we assumed $p_0>0$) by the time it reaches $t=T$.   The same conclusion would be true about the original trajectory $q(t)$ provided we can show that $q$ and $\tilde{q}$ are uniformly close in the interval $[t_0,T]$.  By a similar argument, any trajectory starting to the left of $q_-(0)$ at time $t=T$ must end up to the left of $Q_-$, will be (almost) entirely guided by $\phi_+$, etc., and will have \edit{an approximately} constant velocity $-v_0<0$ by the time it reaches $t=T$.
In order to complete the proof of the Theorem, we only need to show that the actual guiding velocity 
$v^{(\psi)}$
and the one coming from the stationary phase approximation $v^{(U)}$ are sufficiently close, so that the trajectories guided by them are uniformly close for $t \in [t_0,T]$.
\begin{lemma}
Given $\epsilon_0>0$, $p_0 \ne 0$, $T>0$, and $\sigma>0$ there exists $\Omega_0 = \Omega_0(\epsilon_0,p_0,T,\sigma)>0$ such that if $\omega > \Omega_0$, for the trajectories $s=q(t)$ and $s=\tilde{q}(t)$ defined in the above, we have
\begin{equation}
    |q(t) - \tilde{q}(t)| < \epsilon_0\qquad \forall t \in [t_1,T].
\end{equation}
\end{lemma}
\begin{proof}
First we note that for all $t\in [t_1,T]$ and for any $s\in (-t-A\sigma, t+A\sigma)$ we have 
$$|\phi_\pm(t,s)|^2 = f(s\pm v_0 t)^2 = \frac{e^{-(s\pm v_0 t)^2/2\sigma^2}}{\sqrt{2\pi\sigma^2}} >  \frac{e^{-((1+v_0) T +A\sigma)^2/2\sigma^2}}{\sqrt{2\pi\sigma^2}} := \delta>0,
$$
while clearly $|\phi_\pm|^2 \leq \frac{1}{\sqrt{2\pi\sigma^2}} := C^2$.

Let $V_\pm := \psi_\pm - U_\pm$. From the error estimate for SPA we know that for any given $\epsilon>0$, $|V_\pm(t,s)| \leq \epsilon$ for all $t\in [t_0,T]$ and $s\in \mathbb{R}$ provided $\omega>\Omega_0(\epsilon,p_0,T,\sigma)$.  It is then easy to see that
\begin{eqnarray*}
    |v^{(\psi)} - v^{(U)}|(t,s) &\leq & \frac{2 \epsilon^2 + 2\sqrt{2}\epsilon \sqrt{|U_-|^2+|U_+|^2}}{|U_-+V_-|^2+|U_++V_+|^2}\\
    & \leq & \frac{2\epsilon^2 + 2 \sqrt{2} \epsilon\sqrt{f_-^2 + f_+^2}}{\cos^2\frac{\Theta_0}{2} f_-^2 + \sin^2\frac{\Theta_0}{2}f_+^2 - 2\sqrt{2} \epsilon \sqrt{f_-^2+f_+^2}},
\end{eqnarray*}
where $f_\pm := f(s\pm v_0 t)$.  For $ t \in [t_1,T]$ and $s\in (-t-A\sigma, t+A\sigma)$ we thus have $\delta \leq f_\pm^2\leq C^2$.  In that case
\begin{equation}
    |v^{(\psi)} - v^{(U)}|(t,s) \leq \frac{2 (\epsilon+2C)}{\delta - 4 C \epsilon}\epsilon < \frac{\epsilon_0}{T-t_1},
\end{equation}
if $\epsilon \in (0,\frac{\delta}{8C})$ is chosen sufficiently small. 

Next we observe 
\begin{eqnarray}
\left|\frac{d}{dt}(q(t) - \tilde{q}(t))\right| & =  & |v^{(\psi)}(t,q(t)) - v^{(U)}(t,\tilde{q}(t))| \nonumber\\
&\leq &  |v^{(\psi)}(t,q(t)) -v^{(U)}(t,q(t))| + |v^{(U)}(t,q(t)) - v^{(U)}(t,\tilde{q}(t))| \nonumber\\
& \leq & \frac{\epsilon_0}{T-t_1} + M \ |q(t) - \tilde{q}(t)|, \label{estvel}
\end{eqnarray}
where $M>0$ is a constant such that
$$ M \geq \sup_{t \in [t_1,T],s\in [-t-A\sigma,t+A\sigma]}|\partial_s v^{(U)}(t,s)| $$
(it is easy to see that the supremum is finite.)  Integrating \eqref{estvel} on $[t_1,t]$ we obtain
\begin{equation}
    q(t) - \tilde{q}(t) = \int_{t_1}^t \frac{d}{dt'}(q(t') - \tilde{q}(t')) dt' \leq \epsilon_0 + M\int_{t_1}^t |q(t') - \tilde{q}(t')| dt',
\end{equation}
and thus by Gronwall's inequality, $|q(t) - \tilde{q}(t)| \leq \epsilon_0 e^{M(T-t_1)}$, which can be made arbitrarily small.
\end{proof}
\section{Summary and Outlook}
In this paper we studied the relationship between the Bohmian momentum \eqref{def:vpE} and the expected value of the momentum operator \eqref{eq:ExpValP} for solutions of the Dirac equation \eqref{eq:DirEl} with Gaussian wave packet initial data of the form \eqref{Gaussiandata}, for which the latter quantity is $k_0$.  The initial data depends on parameters $\Theta_0,\Omega_0,\sigma$, and $k_0$, while another parameter $\omega := \frac{mc}{\hbar}$, appears in the equation\edit{.} 
We derived the stationary phase approximation for the solution and showed that it corresponds to two wave packets with the same momentum and opposite values of energy that travel in opposite directions, and thus their supports become almost disjoint after enough time has passed. By analyzing the guiding equation \eqref{eq:guiding} we showed that as a result, particles will eventually be influenced only by one of these asymptotic wave packets, so that their momentum becomes \edit{approximately} constant.
\edit{As a result, }the wave function in effect becomes \edit{locally like a plane wave } after enough time has passed, which provides some justification for Compton's use of the plane wave dispersion relations in his calculations, at least as far as the incoming electron is concerned.  Doing a similar analysis for the incoming photon is hampered by the Bohmian momentum being ill-defined for a massless particle, but we believe similar results can be obtained through a limiting argument.  Meanwhile, justifying Compton's use of plane-wave values for the {\em outgoing} particles requires addressing the entanglement that would result from their interaction.  We leave these issues for future works. 
\appendix
\section{Proof of Stationary Phase Approximation Theorem}
\begin{proof}
To establish the SPA, we convert our integrals for $\psi_-$ and $\psi_+$ into integrals on a sphere.
 We use the change of variables $\sigma = s - t\cos\theta$ and substitute for the integral form of the Bessel functions in \eqref{sol:psiminus} and \eqref{sol:psiplus}:
 $$J_n(x) = \frac{1}{2\pi}\int_{-\pi}^{\pi} e^{i(n\varphi - x\sin\varphi)}\,d\varphi.$$

For $\psi_+$ specifically, we first use Lemma \ref{lem:trans} to dispense with the pure transport term. Let $E(t)$ be the error in that approximation.  
With this in mind, the expressions for $\psi_-$ and $\psi_+$ are:
\begin{equation}
    \psi_- = \frac{-i\omega t}{4\pi}\int_{-\pi}^{\pi}\int_0^{\pi} f(s-t\cos\theta)e^{i\omega\left(p_0(s-t\cos\theta)-t\sin\theta\sin\varphi\right)}\,dV
\end{equation}
and
\begin{equation}
    \psi_+ = \frac{-\omega t}{4\pi} \int_{-\pi}^{\pi}\int_{0}^{\theta_0}f(s-t\cos\theta)\frac{1-\cos\theta}{\sin\theta} e^{i\varphi}e^{i\omega(p_0(s-t\cos\theta) - t\sin\theta\sin\varphi)}\,dV + E(t),
\end{equation}
where $dV = \sin\theta\,d\theta\,d\varphi$ is the volume form of the standard unit sphere, and 
$$ 
\theta_0 := \arccos\left({\frac{2\delta}{(\omega t)^2} - 1}\right),\qquad \delta := \frac{j_0^2}{4}.
$$
The assumption $\omega t > j_0/2$ implies that $\theta_0$ is well-defined and $\theta_0 \in (0, \pi)$.  Also note that the integrand for $\psi_+$ is singular at $\theta = \pi$, but not at $\theta = 0$.

Next, since we want to use formulas for integrals over the plane, we use the stereographic projection to write these as integrals over the plane. We use these substitutions: 
\begin{equation}
\label{cov}
\theta = 2\arctan(\frac{1}{\sqrt{x^2+y^2}}), \hspace{0.5 cm} \varphi = \arctan(\frac{y}{x}),
\end{equation}
which imply $$dV = \frac{4}{(1+x^2 + y^2)^2} \,dx\,dy.$$
With these substitutions, and defining
$$
r(x,y) := \sqrt{x^2+y^2},
$$
we find the new expressions for $\psi_-$ and $\psi_+$ to be
\begin{equation}\label{newpsim}
    \psi_- = \frac{-i\omega t}{4\pi}\int\int_{\mathbb{R}^2 }f(s-t\frac{r^2-1}{r^2+1})\frac{4}{(r^2+1)^2}e^{i\omega\left(p_0(s-t\frac{r^2-1}{r^2+1})-t\frac{2y}{r^2+1}\right)}\,dx\,dy
\end{equation}
and
\begin{equation}\label{newpsip}
    \psi_+ = \frac{-\omega t}{4\pi} \int\int_{r\geq R} f(s-t\frac{r^2-1}{r^2+1})\frac{4}{r(1+r^2)^2} e^{i\varphi}e^{i\omega\left(p_0(s-t\frac{r^2-1}{r^2+1})-t\frac{2y}{r^2+1}\right)}\,dx\,dy.
\end{equation}
(We have dropped the error term $E(t)$ in \eqref{newpsip} since according to Lemma~\ref{lem:trans} this error will be of a higher order in $\omega$, and thus much smaller than the ultimate error for the SPA given in Theorem~\ref{thm:main}.)
For $\psi_-$, the domain of integration is $\mathbb{R}^2$, while the domain for $\psi_+$ is the region in the $xy$-plane {\em outside} the circle of radius 
$R = \frac{\sqrt{\delta}}{\sqrt{\omega^2t^2-\delta}}$ centered at the origin.  

We can see that our phase function is 
\begin{equation}\label{phasefun}
\phi(x,y) := p_0(s-t\frac{x^2+y^2-1}{x^2+y^2+1})-t\frac{2y}{x^2+y^2+1} = p_0s - t \frac{p_0 (r^2 -1) + 2y}{r^2+1}.
\end{equation}
(Note that $s,t,p_0$ appear in the phase function as parameters.)
Its gradient is
$$\nabla \phi = -t \nabla \left(\frac{p_0 (r^2 -1) + 2y}{r^2+1}\right) = \frac{-2t}{(r^2+1)^2}\left(2(p_0-y)x\ ,\ x^2 - y^2+2p_0 y + 1\right).$$
From here the Hessian of the phase function is: $$\frac{-4t}{(r^2+1)^3}\begin{bmatrix}
    (p_0-y)(-3x^2+y^2 + 1) & -x(x^2 - 3y^2 +1 + 4 p_0 y) \\ -x(x^2 - 3y^2 +1 + 4 p_0 y) & (p_0-y)(x^2+y^2+1) -2y(x^2 - y^2 +2 p_0 y +1) \\
\end{bmatrix}.$$
We can see from this that the phase function has two critical points $(x=0,y=y_\pm)$ with
$$y_\pm := p_0 \pm \sqrt{p_0^2+1}.$$
We observe that $y_+>0$ and $y_-<0$ regardless of the sign of $p_0$.

Next we evaluate the Hessian of the phase function at the two critical points $(0,y_\pm)$, to obtain 
$$D^2\phi(0,y_\pm) = \frac{\pm t}{\sqrt{p_0^2+1}(p_0 \pm \sqrt{p_0^2 + 1})^2}\begin{bmatrix}1 & 0 \\ 0 & 1  \\ \end{bmatrix}.$$
Accordingly, the signature (number of positive eigenvalues minus number of negative eigenvalues) of the Hessian of $\phi$ is equal to $2$ at $(0,y_+)$ and $-2$ at $(0,y_-)$.  

The well-known formula for the stationary phase approximation of an oscillatory integral over an open domain $D \subset \mathbb{R}^2$ of the form,
\begin{equation}\label{oscint}
I(\omega) := \iint_D g(x) e^{i \omega \phi(x)} d^2 x,    
\end{equation} 
with amplitude function $g$ and phase function $\phi$ is as follows:  If $x_0$ is the only critical point of $\phi$ inside $D$, then, as $\omega \to \infty$, the leading order contribution to $I(\omega)$ is
\begin{equation}   \label{SPA}
U_1(\omega) = g(x_0) e^{i \omega \phi(x_0)} \frac{e^{i\ \mbox{sgn}(\mbox{Hess}(x_0)) \pi/4}}{\sqrt{|\det(\mbox{Hess}(x_0))|}} \frac{2\pi}{\omega}, 
\end{equation}
while if there is more than one critical point in $D$, the stationary phase approximation is simply the sum of contributions as above from each critical point.  There is also an extension to critical points on the boundary of $D$, where it is shown that the contribution of those is half the value above. 

In our case by \eqref{newpsim} and \eqref{newpsip} we have two such oscillatory integrals
$$
\psi_\pm = \frac{-\omega t}{4\pi} \iint_{D_\pm} g_\pm(x,y) e^{i \omega \phi(x,y)} dx\,dy =: \frac{-\omega t}{4\pi} I^\pm(\omega),
$$
where $D_- = \mathbb{R}^2$ and $D_+ = \mathbb{R}^2 \setminus B_R(0)$, $\phi$ is as in \eqref{phasefun}, and
\begin{eqnarray}
    g_\pm(x,y) & := & f(s - t \frac{r^2 - 1}{r^2+1}) \tilde{g}_\pm(x,y)\\
    \tilde{g}_-(x,y) & := & 4i (1+r^2)^{-2} \\
    \tilde{g}_+(x,y) & := & 4 r^{-1} (1+r^2)^{-2} e^{i\varphi}.
\end{eqnarray}
By the change of coordinates formula \eqref{cov} we have that if the point $(\theta_\pm,\varphi_\pm)$ corresponds to the critical point $(0,y_\pm)$, then 
$$
\varphi_\pm = \pm \frac{\pi}{2},\qquad \cos\theta_\pm = \frac{y_\pm^2 - 1}{y_\pm^2 +1} = \frac{\pm p_0}{\sqrt{1+p_0^2}}.
$$
Recalling that the largest value of $\theta$ that would be inside the domain $D_+$ is $\theta = \theta_0$ with $\cos \theta_0 = -1 + \frac{2\delta}{(\omega t)^2}\leq -1/2$, it follows that regardless of the sign of $p_0$, one of the critical points, the one with $\cos\theta>0$ will always be in the domain, while the one with $\cos\theta<0$ will only be included if $\omega t$ is sufficiently large.  In particular it would suffice that
$$ \omega t > j_0 \sqrt{1+p_0^2},$$
for the other critical point to also be included in $D_+$.  Thus the SPA for $\psi_+$ is different depending on whether or not $\omega t$ is in this range. 

Consider first the case when both critical points are included in $D_+$.  Using \eqref{SPA} repeatedly, a calculation then shows
\begin{equation}
    U^-_1  = \frac{e^{i\omega p_0 s}}{2\sqrt{1+p_0^2}}\left[f(s-v_0t)e^{-i\omega t\sqrt{1+p_0^2}} - f(s+v_0t)e^{i\omega t\sqrt{1+p_0^2}}\right],
\end{equation}
with $v_0 := p_0/\sqrt{1+p_0^2}$, while
 \begin{equation}\label{Uplus1}
    U^+_1 = \frac{e^{i\omega p_0 s}}{2\sqrt{1+p_0^2}}\left[f(s-v_0t)(\sqrt{1+p_0^2}-p_0)e^{- i\omega t\sqrt{1 + p_0^2}} + f(s+v_0t)(\sqrt{1+p_0^2}+p_0)e^{i\omega t\sqrt{1 + p_0^2}} \right].
\end{equation}
If, on the other hand, one has $j_0/2 < \omega t < j_0 \sqrt{1+p_0^2}$, then only one of the critical points, the one with smaller angle $\theta$ will be included, as a result of which \eqref{Uplus1} has to be modified by keeping only the contribution of that critical point.

This completes the derivations of the SPA expressions for $\psi_\pm$.  Next we must examine the error associated with these approximations. In doing so we will rely on the approach of Chako \cite{Chako}, which we briefly outline here:
Consider an oscillatory integral of the form \eqref{oscint}, and let $(x_0,y_0)$ be a critical point of $\phi$ inside $D$.  Assume that the amplitude and phase functions have the following expansions at this critical point:
\begin{eqnarray}
    \phi(x,y) & = & \phi(x_0,y_0) + a_{\sigma,0} (x-x_0)^\sigma [1+ P(x,y)] + b_{0,\tau} (y-y_0)^\tau [1+Q(x,y)]\label{phiexp}\\
    g(x,y) & = & (x-x_0)^{\lambda_0 - 1}(y-y_0)^{\mu_0-1} g_1(x,y)\\
    P(x,y) & = & \sum_{m+n\geq 1} p_{m,n} (x-x_0)^m (y-y_0)^n \label{psP}\\
    Q(x,y) & = & \sum_{m+n\geq 1} q_{m,n} (x-x_0)^m (y-y_0)^n \label{psQ}\\
    g_1(x,y) & = & \sum_{k+\ell\geq 0} g_{k\ell} (x-x_0)^k (y-y_0)^\ell.\label{psg}
\end{eqnarray}
Let $\alpha,\beta>0$ be chosen small enough such that the rectangle 
$$ R := \{(x,y)\in D\ |\ |x-x_0|<\alpha, |y-y_0|<\beta\}$$
lies within the radius of convergence of all of the power series (\ref{psP},\ref{psQ},\ref{psg}).  Let $C_{PQ},C_g>0$ be constants such that
\begin{equation}\label{trivest}
|g_{k\ell}| < \frac{C_g}{\alpha^k \beta^\ell}, \forall k,\ell\geq 0,\qquad |p_{m,n}|,|q_{m,n}| < \frac{C_{PQ}}{m \alpha^m n \beta^n}, \forall m,n\geq 1.    
\end{equation}
(The existence of these constants is guaranteed by the absolute convergence of the power series within their radius of convergence.)  Chako then uses standard majorization and contour integration techniques to estimate the remainder $R_{1,1}$ in \eqref{UandR} in the following way.  He proves that there exists a constant $K>0$ such that
\begin{eqnarray}
    |R_{1,1}(\omega)| & \leq & K C_g (\omega a_{\sigma,0})^{-\lambda_0/\sigma} (\omega b_{0,\tau})^{-\mu_0/\tau} \\
    &\times& \!\!\!\!\!\int_0^\infty\!\!\!\! \int_0^\infty \!\!\!\! e^{-u/2} e^{-v/2} u^{\lambda_0/\sigma} v^{\mu_0/\tau} \left[ \frac{(1+C_{PQ} u)}{\alpha} \left(\frac{u}{\omega a_{\sigma,0}}\right)^{1/\sigma}\!\!\!\!\!\! + \frac{(1+C_{PQ} v)}{\beta} \left(\frac{v}{\omega b_{0,\tau}}\right)^{1/\tau}\right] du dv.\nonumber
\end{eqnarray}
Evaluating the resulting Gamma-function-type integrals, he then obtains
\begin{equation}\label{ChakoRem}
    |R_{1,1}(\omega)| \leq C_{1,1} \omega^{-(\lambda_0/\sigma+\mu_0/\tau)} \left[ O(\omega^{-1/\sigma}) + O(\omega^{-1/\tau})\right],
\end{equation}
for some constant $C_{1,1}$ that depends on all the other aforementioned constants.  In our case, $\lambda_0 = \mu_0 = 1$, $\sigma = \tau = 2$, and $a_{2, 0} = b_{2, 0} = \frac{\pm t}{2\sqrt{p_0^2 + 1}( p_0 \pm \sqrt{p_0^2 + 1})^2}$.  The main difficulty in using Chako's approach for us is that our phase and amplitude functions depend on space variable $s$ and time variable $t$, as well as initial data parameters $p_0$, and $\sigma$, and we need to know how this dependence is reflected in the constants $C_g$ and $C_{PQ}$. (Chako does not even distinguish between these two constants, but in our case they are very different, as we shall see below.)

We begin by recalling standard formulas for generalized Leibnitz rule and generalized chain rule (Faa di Bruno rule.) Let $\alpha = (\alpha_1,\dots,\alpha_n)$ and $\beta = (\beta_1,\dots,\beta_n)$ denote two multiindices, and suppose $u$ and $v$ are two functions defined on $\mathbb{R}^n$.  Then
\begin{equation}\label{genLeib}
    \partial^\alpha (uv) = \sum_{\beta \leq \alpha} \begin{pmatrix} \alpha \\ \beta \end{pmatrix} \partial^\beta u \partial^{\alpha - \beta} v.
\end{equation}
Suppose further that $h :\mathbb{R}^n \to \mathbb{R}$ and $f : \mathbb{R}\to \mathbb{R}$.  Then
\begin{equation}\label{genchain}
    \partial^\beta (f\circ h) = \sum_{\pi \in \Pi_\beta} f^{(|\pi|)}(h) \Pi_{B\in \pi} \frac{\partial^{|B|} h}{\Pi_{j\in B} \partial x_j},
\end{equation}
where $\Pi_\beta$ denotes the set of all partitions of the multi-index $\beta$. 

We now recall our phase and amplitude functions, rewriting them slightly to make the dependence on various parameters easier to deal with. Recall $r = \sqrt{x^2+y^2}$ and $\varphi = \arctan(y/x)$. We have
\begin{eqnarray}
    \phi(x,y; s,t,p_0) & = & p_0s - t \tilde{\phi}(x,y;p_0)\label{phi}\\
    \tilde{\phi}(x,y;p_0) & := & p_0 + 2 \frac{y-p_0}{r^2+1}\\
    g_{\pm}(x,y; s,t,\sigma) & = & f(h(x,y;s,t))\tilde{g}_\pm(x,y)\label{gpm}\\
    h(x,y;s,t) & := & s - t \tilde{h}(x,y) \\
    \tilde{h}(x,y) & := & 1 - 2(r^2+1)^{-1} \\
    \tilde{g}_-(x,y) & := & 4 i (r^2+1)^{-2} \\
    \tilde{g}_+(x,y) & := & 4 r^{-1} (r^2+1)^{-2} e^{i\varphi}. \label{gplus}
\end{eqnarray}
We are interested in the Taylor expansions of these functions that are valid in a small neighborhood of the two critical points $(0,y_\pm)$ of $\phi$.  Since $y_\pm$ depends on $p_0$, we need to do an appropriate scaling so that the dependence of the Taylor coefficients on $p_0$ become explicit.  We first note that if $p_0>0$, then $y_+ \approx 2p_0$ and $y_- \approx \frac{-1}{2p_0}$, and this is reversed for $p_0<0$, i.e. $y_+ \approx \frac{-1}{2p_0}$ and $y_- \approx 2p_0$. 

Consider now the expansion of $\tilde{h}(x,y)$ centered at $(0,y_\pm)$:  Letting $\tilde{y} := y -y_\pm$, we have
$$
\tilde{h}(x,y) = 1- \frac{ 2}{x^2 + \tilde{y}^2 + 2 y_\pm \tilde{y} + y_\pm^2+1} = 1 - \frac{2}{u^2}\frac{1}{\xi^2 + \eta^2 + 2 a \eta  + b } =: 1 - \frac{2}{u^2} H(\xi,\eta),
$$
where we have introduced scaled variables 
$$\xi := \frac{x}{u},\qquad \eta := \frac{\tilde{y}}{u}, \qquad a := \frac{y_\pm}{u}, \qquad b := \frac{y_\pm^2+1}{u^2}.
$$
Thus, choosing the scaling factor $u$ appropriately, i.e. either $u=2p_0$ or $u=1/2p_0$, we can make sure that $a$ and $b$ are always $O(1)$ as $|p_0|$, which by assumption is bounded away from zero, becomes large, so any derivative of $\tilde{h}$ at $(0,y_\pm)$ is going to be bounded above in the worst case by $p_0^2$ times the corresponding derivative of the function $H$ at $(0,0)$.  The coefficients in the Taylor expansion of $H$ will satisfy estimates of the type in \eqref{trivest} trivially, and the corresponding coefficients for $\tilde{h}$ will at most be a factor of $p_0^2$ larger.  It is not hard to see that analogous statements hold for the Taylor coefficients of $\tilde{\phi}$ and $\tilde{g_\pm}$, but with a different power of $p_0$ (matching the power of $r$ in their denominators).  The highest such power is 5 and occurs in $\tilde{g}_+$, see \eqref{gplus}, and that explains the $p_0^5$ factor in the error bound.

Next we need to address the $s$ and $t$ dependence of $C_g$ and $C_{PQ}$.  Substituting \eqref{phi} in \eqref{phiexp} we see that 
$P$ and $Q$ are independent of both $s$ and $t$, so that $C_{PQ}$ will only depend on $p_0$, in the manner described above.  By contrast, from \eqref{gpm} it is evident that the coefficients of the Taylor expansions of $g_\pm$ will certainly have $t$-dependence in addition to the $p_0$ dependence already discussed.  Indeed, using the generalized Leibnitz \eqref{genLeib} and chain \eqref{genchain} rules, we can see that the coefficients in the expansions of $g_\pm$ will feature powers of $t/\sigma$, and thus the only expression that can bound all of them (as is stipulated for $C_g$ in \eqref{trivest}) would be an exponential in $t/\sigma$. 

This establishes the form of the error term \eqref{errorest}.  To see that the error can be made arbitrarily small, let $\epsilon>0$ be given.  Set 
$$
\Omega_0(\epsilon, p_0,T,\sigma) = \max\left\{ \frac{p_0^{10} A^2 e^{2BT/\sigma}}{\epsilon^2}, \frac{2(1+p_0^2)j_0}{|p_0| T}.\right\}
$$
It then follows from \eqref{errorest} that for all $p_0\ne 0$, $T>0$, $\sigma>0$, $s\in \mathbb{R}$, and for  all $\omega > \Omega_0$ and all $t$ in $[\frac{|v_0|}{2}T,T]$, we have $|R_{1,1}(t,s)| < \epsilon$ as desired. 

The above choice for $\Omega_0$ has the added feature that $\Omega_0 T |v_0|/2 > j_0 E_0$, and therefore, at any time $t \in [\frac{|v_0|}{2}T, T]$ and for any $\omega>\Omega_0$, one has $\omega t > j_0 E_0$, and thus the stationary phase approximation $U_1^+$ that's in effect is \eqref{Uplus1}, i.e. the one where both critical points of the phase are included in the domain.
\end{proof} 
 \section{Numerical Investigations}
\begin{figure}
    \centering
    \includegraphics[scale = 0.5]{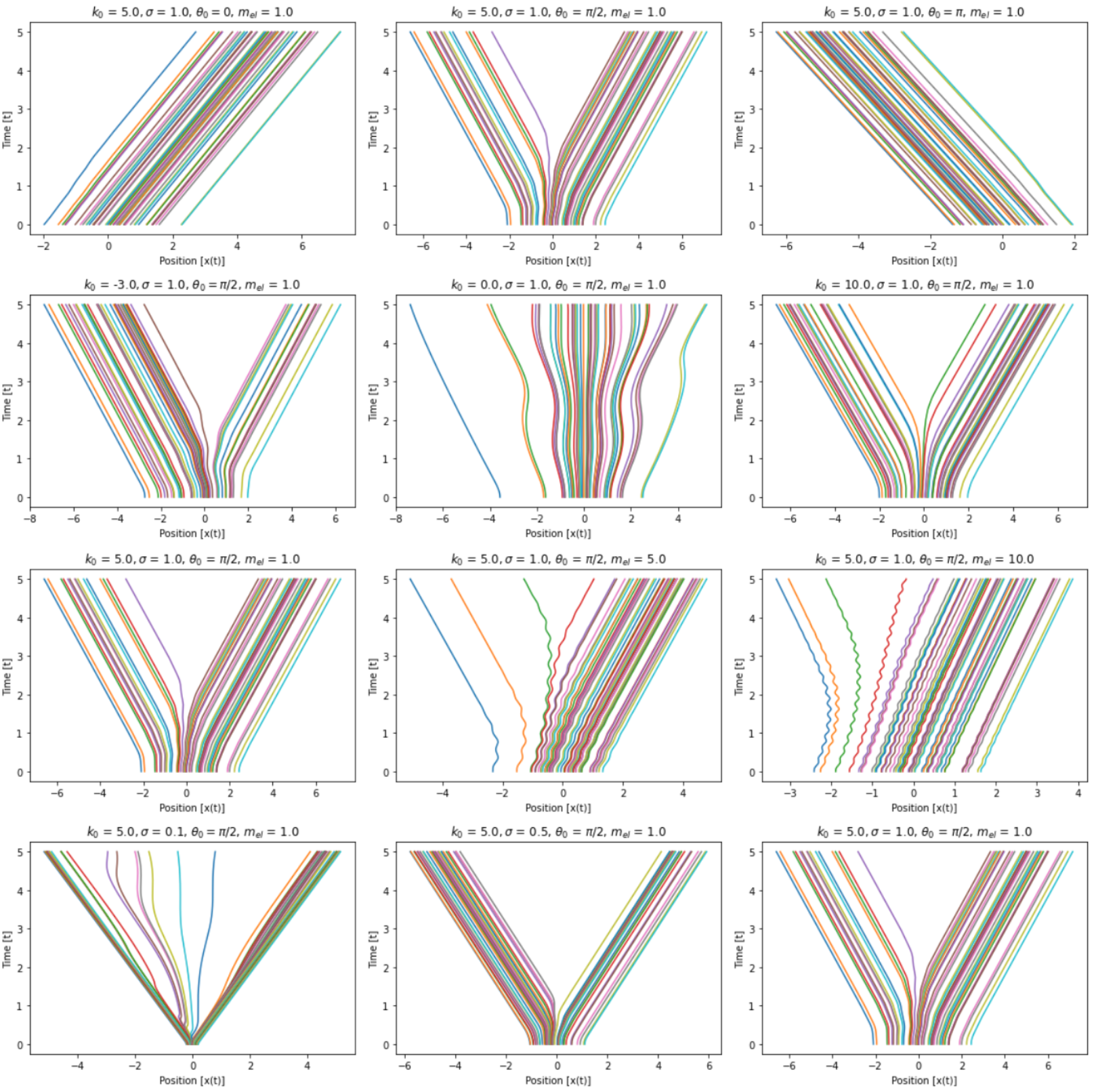}
    \caption{50 electron trajectories. Each row demonstrates the effect of varying one of the initial parameters $\Theta_0$, $k_0$, $m$, and $\sigma$. }
    \label{fig:electron}
\end{figure}

As mentioned before, our focus in this paper was the relationship between the Bohmian momentum $p^{(\psi)}$ as defined in \eqref{def:vpE} and the expected value of the momentum operator \eqref{eq:ExpValP} for solutions of the Dirac equation \eqref{eq:DirEl} with Gaussian wave packet initial data of the form \eqref{Gaussiandata}, where the latter quantity is $k_0$.  The data depends on parameters $\Theta_0,\Omega_0,\sigma$, and $k_0$, while another parameter, the mass of the particle $m$, appears in the equation. 
In addition to this theoretical work we also ran a series of numerical experiments in which we found the wave function at time $t$ using formulas \eqref{psipmsol} for a handful of different values of these parameters 
and solved the guiding equation \eqref{eq:guiding} numerically for a large number of trajectories whose initial positions were distributed randomly according to the same normal distribution given by the initial Gaussian data. See Fig.~\ref{fig:electron} for results. It became clear that for large enough $t$ the Bohmian velocity $v^{(\psi)}$ either approaches $v_0 = k_0/\sqrt{k_0^2 + m^2}$ or $-v_0$, depending on the initial position. 
Specifically, given an initial choice of momentum, say $k_0 > 0$, there is a critical value $s_0$ at which  $q(0) > s_0$ implies $v \rightarrow v_0$ and $q(0) < s_0$ implies $v \rightarrow -v_0$. If $k_0 < 0$, this relationship is reversed.

\begin{figure}
    \includegraphics[width=0.33\linewidth]{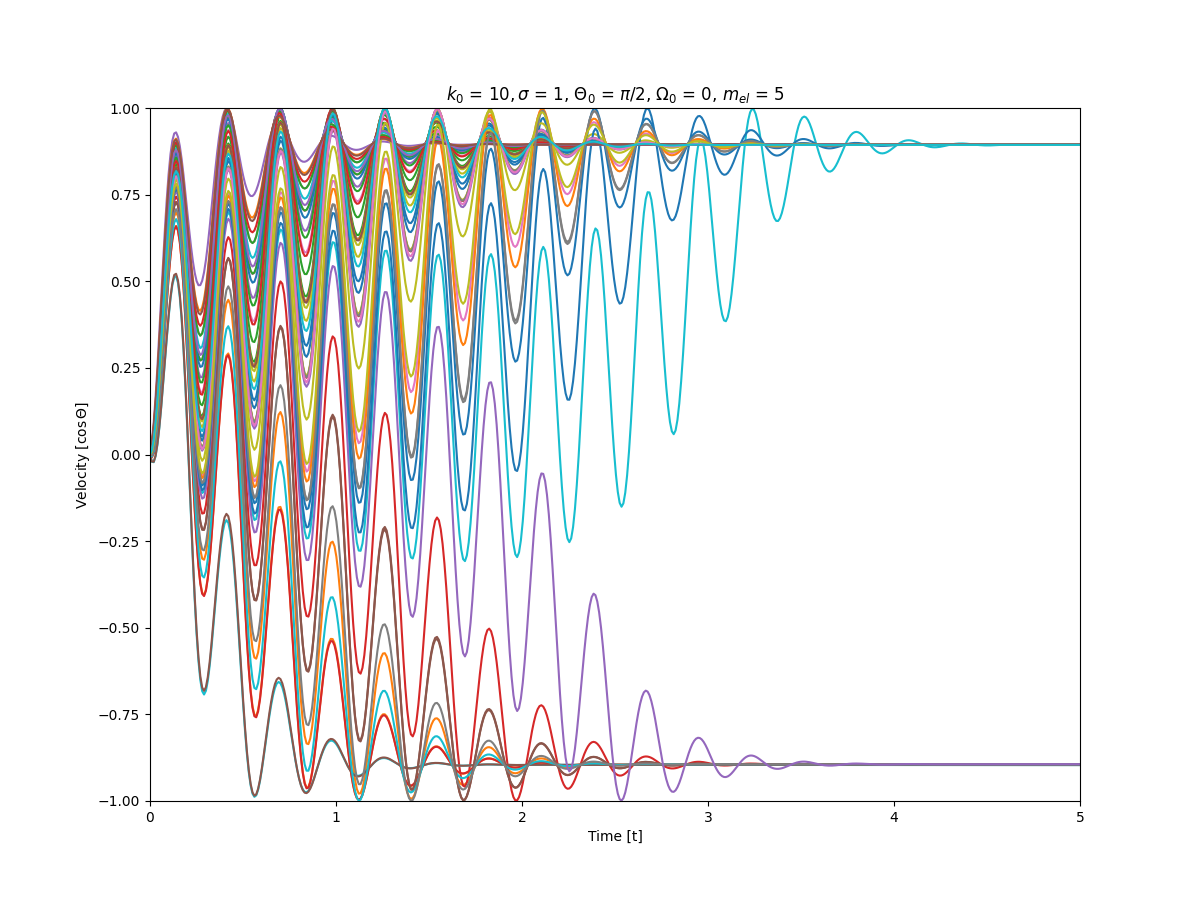}
    \includegraphics[width=0.33\linewidth]{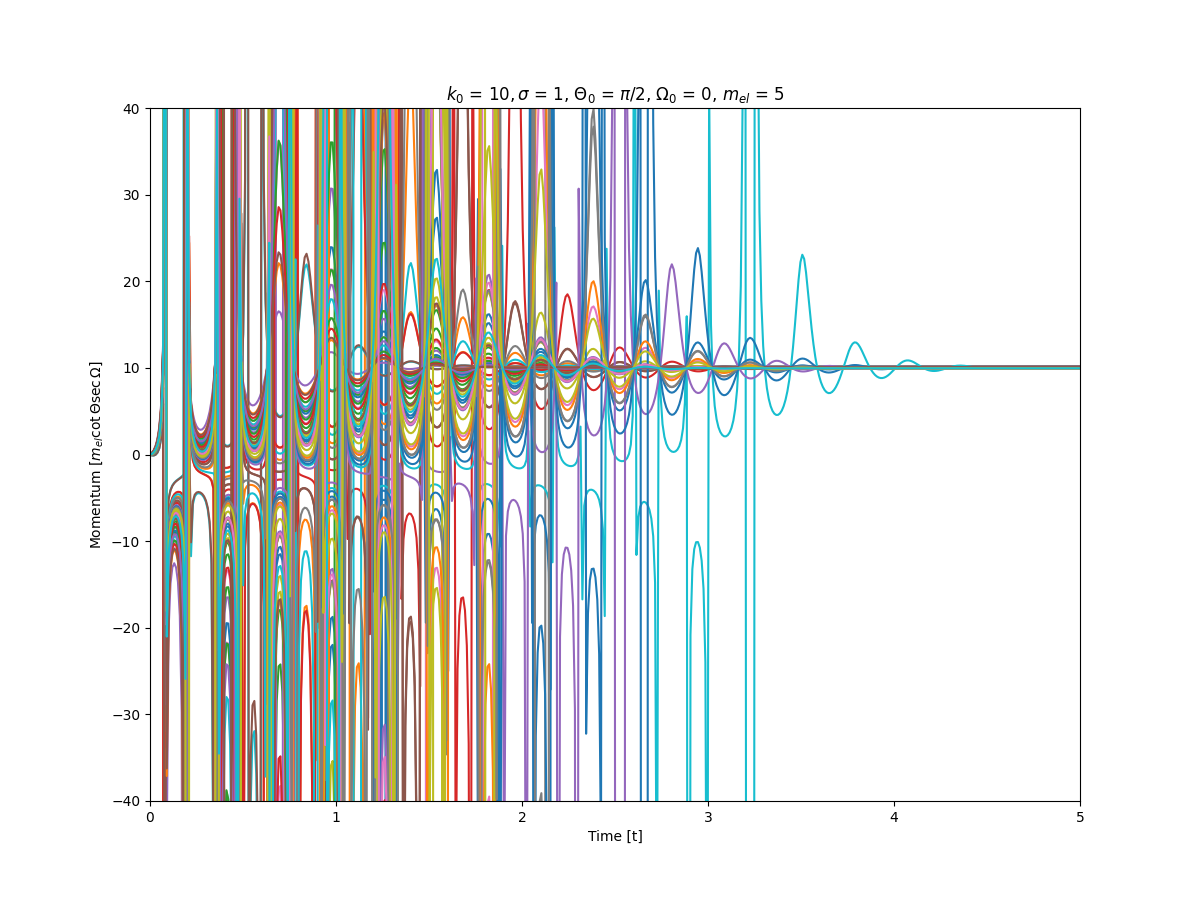}
    \includegraphics[width=0.33\linewidth]{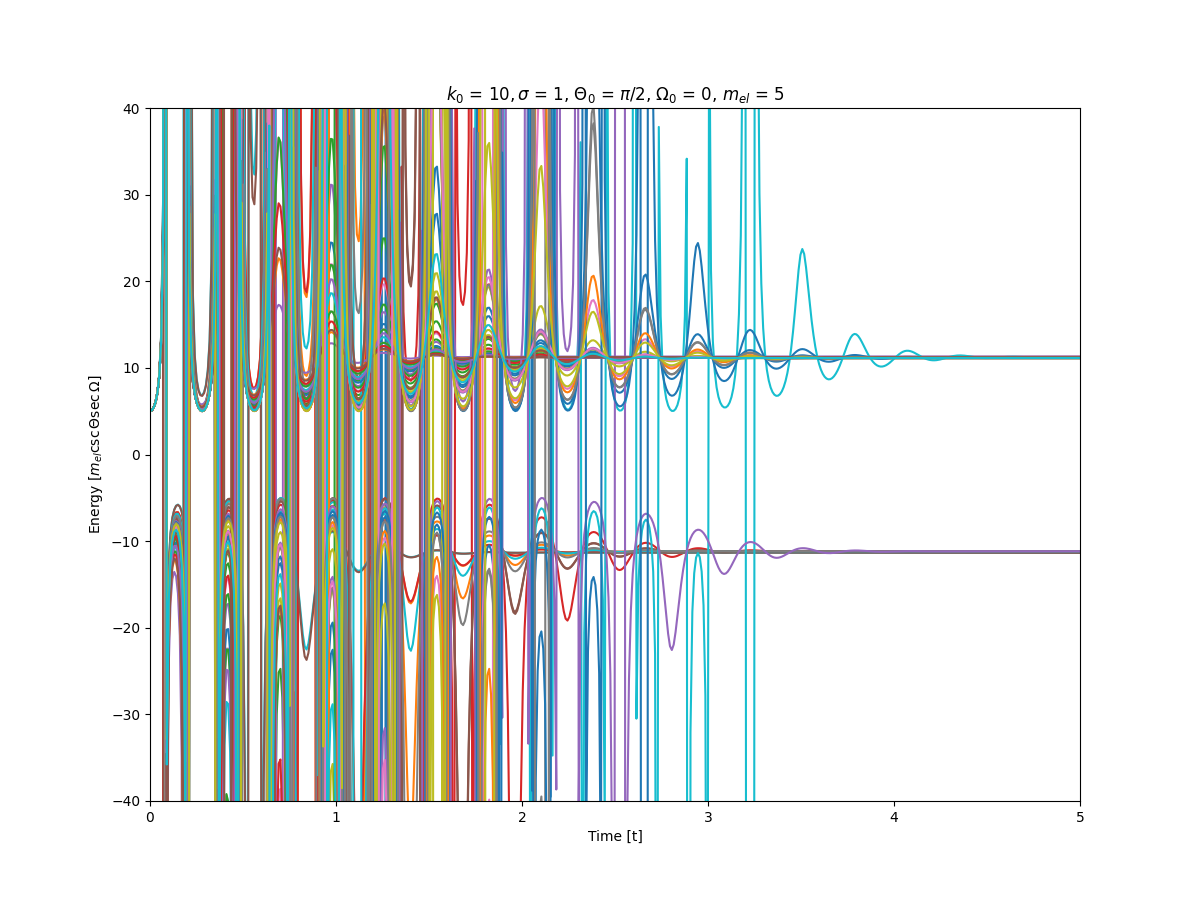}
    \caption{Velocity, momentum, and energy of trajectories with $\Theta_0 = \frac{\pi}{2}$}
    \label{fig:pion2}
\end{figure}

We also verified that the conclusions of Theorem \ref{thm:main} hold for Gaussian wave packet initial data of the form \eqref{Gaussiandata} with $\Theta_0 = \frac{\pi}{2}$ and $\Omega_0= 0$.  We did so by plotting the velocity, momentum, and energy of trajectories with  initial positions distributed according to the corresponding initial probability density. See Fig.~\ref{fig:pion2}.  As expected, there is a bifurcation in velocity and in energy, but not in momentum.

\begin{figure}
    \centering
    \includegraphics[scale = .35]{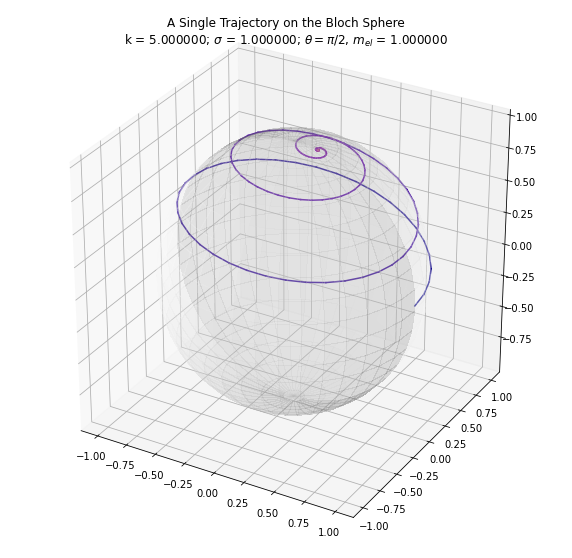}
    \includegraphics[scale = .35]{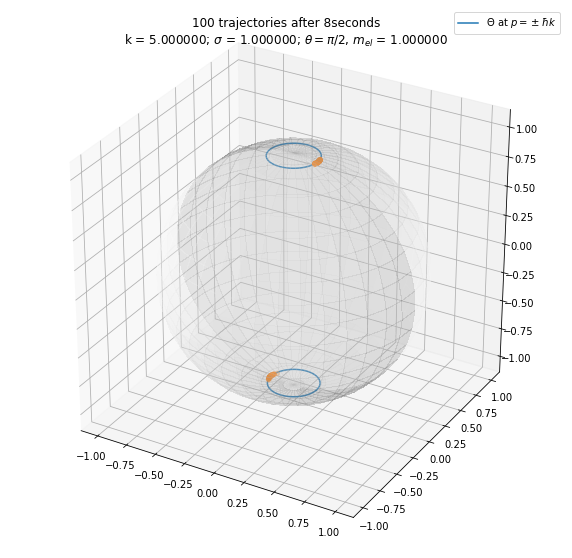}
    \caption{The Bloch sphere representation of a single electron trajectory over 8 seconds (\textit{left}) vs. 100 trajectories after 8 seconds (\textit{right}).}
    \label{fig:bloch}
\end{figure}

\begin{figure}
    \centering
    \includegraphics[scale = .35]{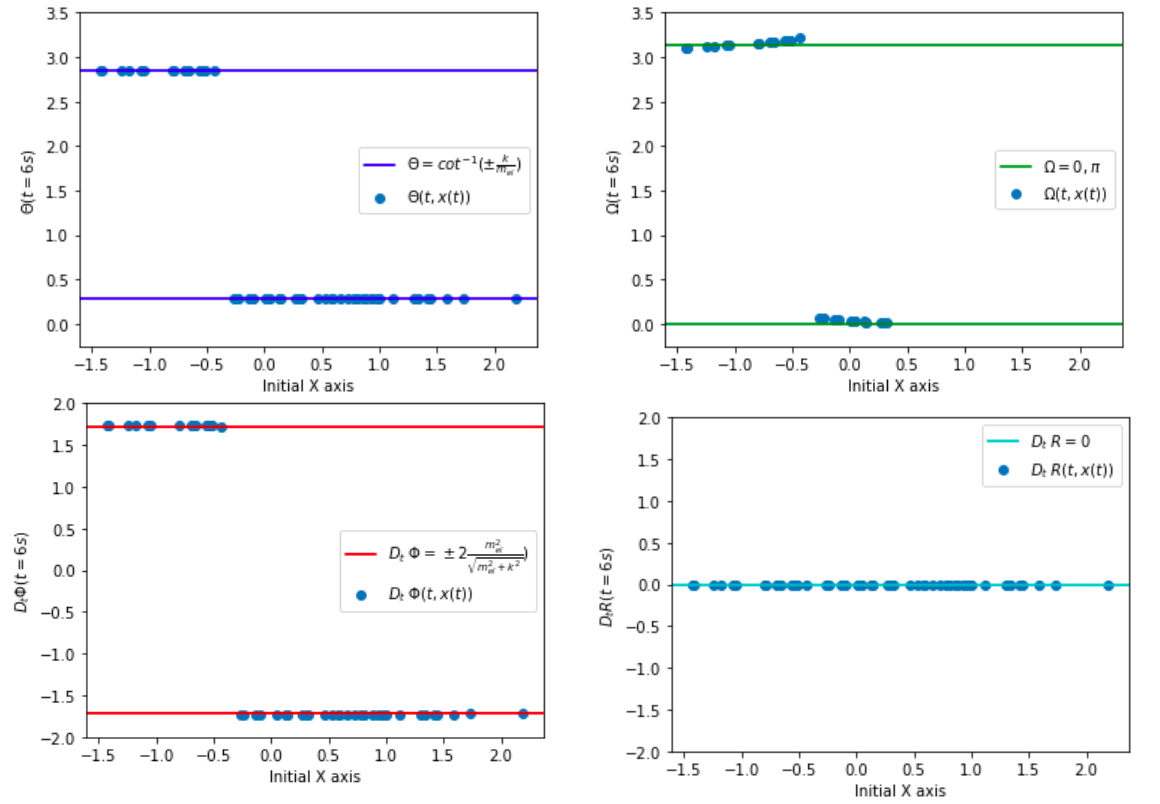}
    \caption{Numerical vs. predicted values of Bloch variables for 50 trajectories after 6 seconds, for $k_0 = 10$, $m = 3$, $\sigma = 1$, $\Theta_0 = \pi/2$.}
    \label{fig:blochvars}
\end{figure}

In conjunction with these results, we examined the electron trajectories in the Bloch sphere representation. 
Notably, the asymptotic behavior aligns exactly with the conclusions in the above. We see each trajectory spiral into one of two antipodal points, as represented in Figure \ref{fig:bloch}. 
We also see that the asymptotic behavior in all four Cayley-Klein parameters align: $R(t,s)$, $\Theta(t, s)$, and $\Omega(t, s)$ all appear to approach constant values, while $\Phi(t, s)$ behaves linearly after some time.
Figure~\ref{fig:blochvars} shows that a bifurcation along the initial $s$-axis in the long run sends trajectories to one or the other side. Where this separation occurs depends on both $k_0$ and $\Theta_0$. At the bifurcation value, the angles $\Theta$ and $\Omega$ jump together, which implies using \eqref{def:vpE} that the asymptotic momentum is always the same as $k_0$ while the sign of the asymptotic energy flips across this divide.   To check the behavior of $\Phi$ and $R$, we plot their time derivatives instead. For $\Phi$ we simply take its total derivative with respect to time, and plug in the asymptotic Bohmian velocity for $\frac{ds}{dt}$ to find the expected asymptotic behavior along a trajectory.

\begin{figure}
    \centering
    \href{https://sites.math.rutgers.edu/~shadit/Bohmian_Momentum/BOMO1.gif}{
        \includegraphics[width=0.3\linewidth]{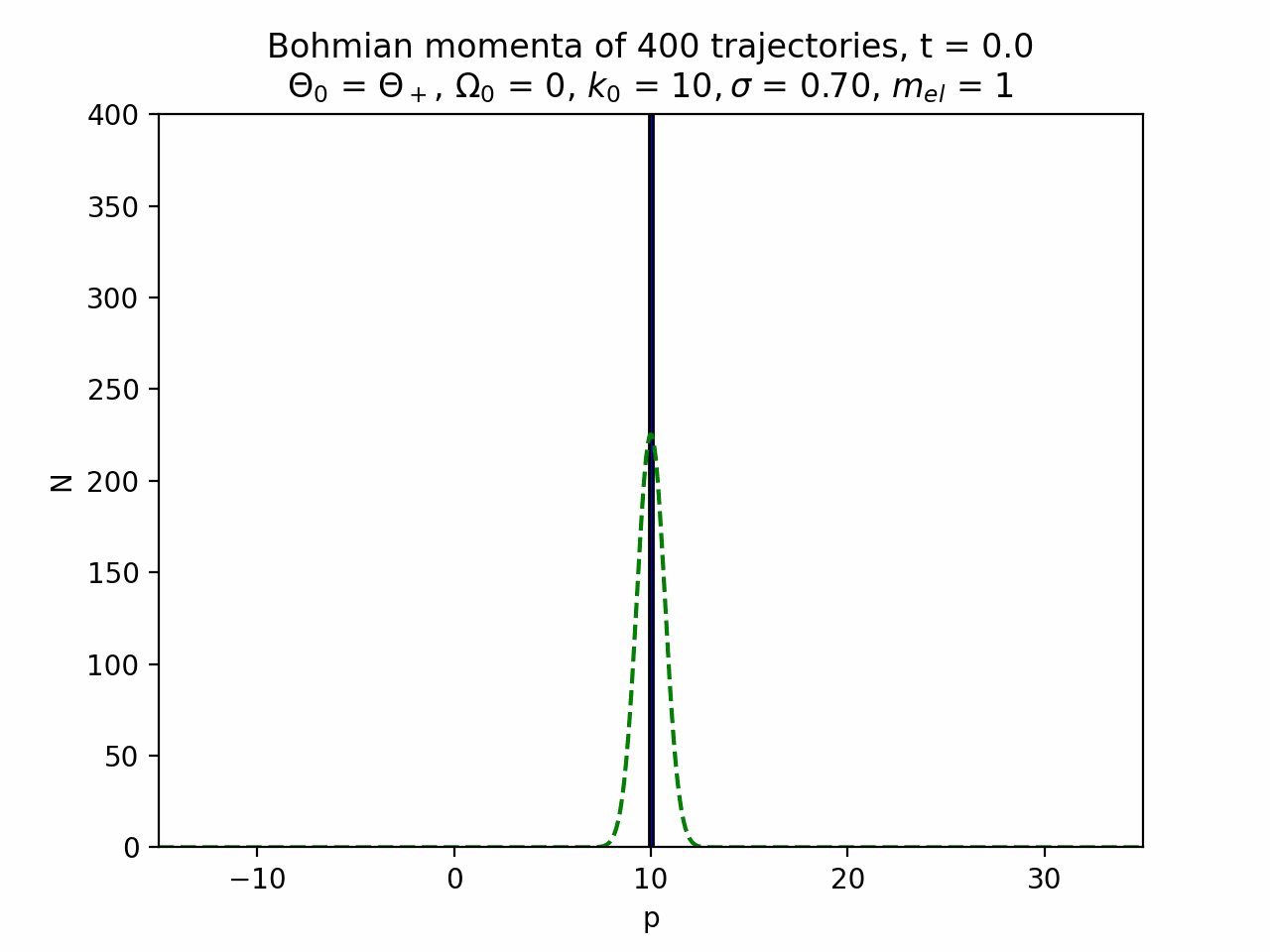}}
    \href{https://sites.math.rutgers.edu/~shadit/Bohmian_Momentum/BOMO2.gif}{
    \includegraphics[width=0.3\linewidth]{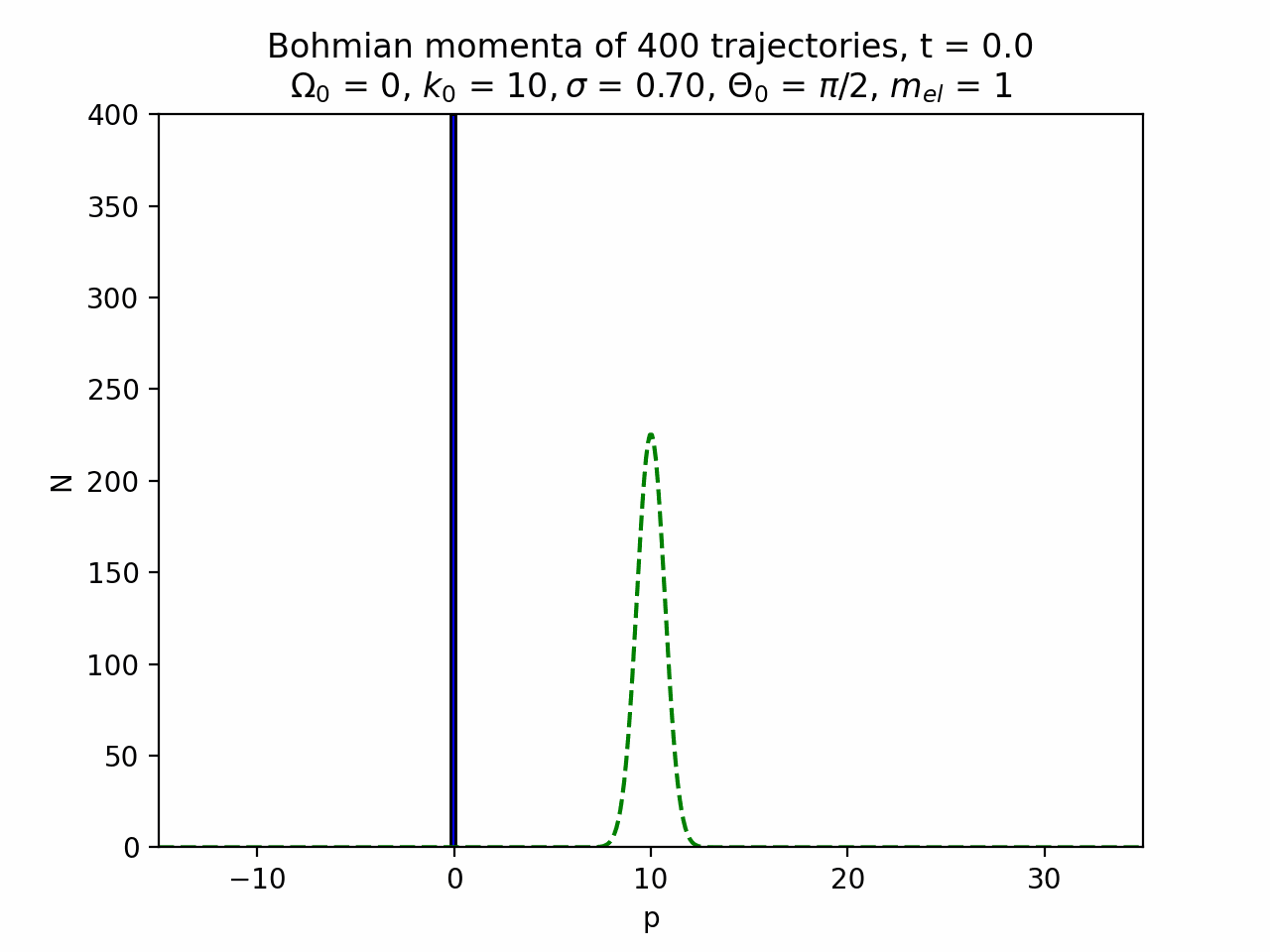}}
    \href{https://sites.math.rutgers.edu/~shadit/Bohmian_Momentum/BOMO3.gif}{
    \includegraphics[width=0.3\linewidth]{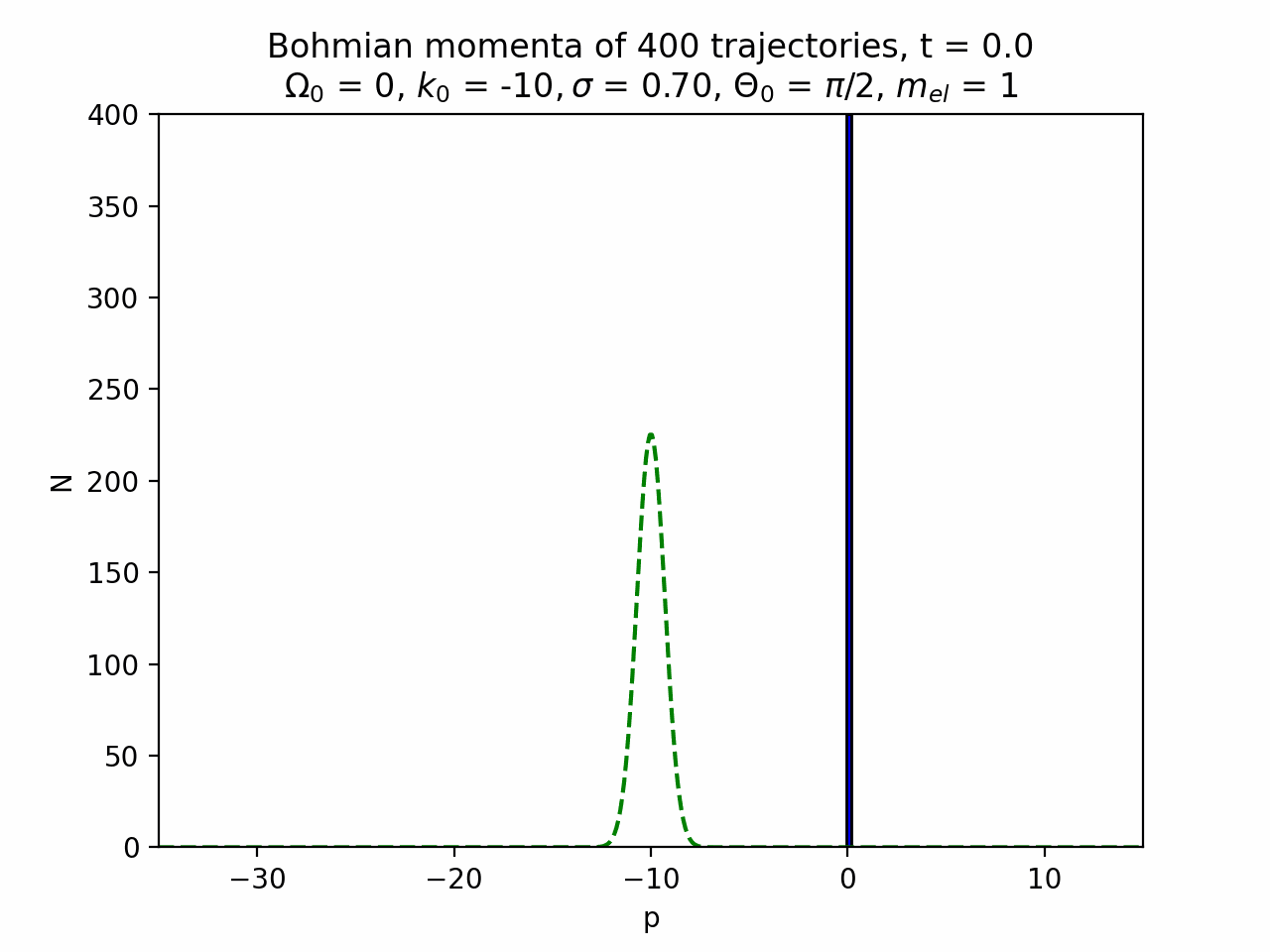}}
    \caption{Evolving Bohmian momentum distributions}
    \label{fig:evolmom}
\end{figure}

\edit{Lastly, we looked at the ``empirical" distribution of the evolving Bohmian momentum along trajectories and compared it to the conserved quantum mechanical momentum distribution of the wave function. The resulting animations can be found \href{https://sites.math.rutgers.edu/~shadit/Bohmian_Momentum/}{here}.
(See Fig.~\ref{fig:evolmom}.)}

\edit{As expected, except for when the initial wave packet is the Gaussian approximation of a positive energy eigenfunction, as in \eqref{posen}, in which case the Bohmian momentum clearly oscillates around the conserved mean before settling down on it, there is little in common between the two distributions for $t$ small. For large $t$ on the other hand, the means appear to converge, while the variance of Bohmian momentum in the long run appears to be much smaller than the conserved variance of the wave function momentum distribution.  It would be interesting to establish these statistical relationships rigorously.}

\bibliographystyle{amsplain}
\bibliography{nptz}
\end{document}